\documentclass[review]{elsarticle}

\usepackage{lineno,hyperref}
\usepackage{color}
\usepackage{amsfonts}
\usepackage{graphics}
\usepackage{graphicx}
\input{epsf}
\usepackage{amsmath}
\usepackage{amssymb}
\modulolinenumbers[5]

\journal{Journal of \LaTeX\ Templates}

\begin{document}

\begin{frontmatter}

\title{Clustering of Excursion Sets in Financial Market}

\author{M. Shadmangohar} 
\address{Department of Physics, Shahid Beheshti University,  1983969411, Tehran, Iran}

\author{S. M. S. Movahed\corref{mycorrespondingauthor}}
\address{Department of Physics, Shahid Beheshti University,  1983969411, Tehran, Iran}
\cortext[mycorrespondingauthor]{Corresponding author}
\ead{m.s.movahed@ipm.ir}


\begin{abstract}
Relying on the excursion set theory, we compute the number density of local extrema and crossing statistics versus the threshold for the stock market indices. Comparing the number density of excursion sets calculated numerically with the theoretical prediction for the Gaussian process confirmed that all data sets used in this paper have a surplus (almost lack) value of local extrema (up-crossing) density at low (high) thresholds almost around the mean value implying universal properties for stock indices. We estimate the clustering of geometrical measures based on the excess probability of finding the pairs of excursion sets, which clarify well statistical coherency between markets located in the same geographical region. 
The cross-correlation of excursion sets between various markets is also considered to construct the matrix of agglomerative hierarchical clustering. Our results demonstrate that the peak statistics is more capable of capturing blocks. Incorporating the partitioning approach, we implement the Singular Value Decomposition on the matrix containing the maximum value of unweighted Two-Point Correlation Function of peaks and up-crossing to compute the similarity measure. Our results support that excursion sets are more sensitive than standard measures to elucidate the existence of {\it a priori} crisis.


\end{abstract}


\end{frontmatter}

\linenumbers

\begin{table}
	\begin{center}
		\scalebox{0.7}{
			\begin{tabular}{|c|c|c|c|c|c|}
				\hline
				Index&Country name & Index&Country name  & Index&Country name   \\\hline
				1&ARGENTINA & 17&HONG-KONG &33&PERU  \\\hline
				2& AUSTRALIA &  18& HUNGARY    &34& POLAND                                                \\\hline
				3&  AUSTRIA &19&INDIA &35& PORTUGAL\\\hline
				4& BELGIUM &20& INDONESIA & 36&RUSSIA \\\hline
				5&  BRAZIL & 21&IRELAND &37&SINGAPORE\\\hline
				6&    CANADA &22& ITALY  &38&SOUTH-AFRICA\\\hline
				7& CHILE & 23&JAPAN &39&SPAIN\\\hline
				8&CHINA &24& JORDAN & 40 & SRI-LANKA\\\hline
				9&COLOMBIA & 25&KOREA & 41&SWEDEN\\\hline
				10&CZECH-REPUBLIC &26&MALAYSIA  & 42&SWITZERLAND \\\hline
				11&DENMARK & 27&MEXICO &43&TAIWAN \\\hline
				12&EGYPT & 28&MOROCCO & 44&THAILAND \\\hline
				13&FINLAND &29& NETHERLANDS& 45&TURKEY  \\\hline
				14&FRANCE& 30&NEW-ZEALAND & 46&UK\\\hline
				15&GREECE & 31&NORWAY& 47&USA \\\hline
				16&GERMANY&32&PAKISTAN&-&- \\\hline
			\end{tabular}}
			\caption{\label{index1}The name of countries and their indices used in this paper.}
		\end{center}
	\end{table}

	\section{Introduction}
	
	Implementation of methods developed in data modeling from the stochastic field point of view has become a topic of interest in econophysics \citep{mantegna1999hierarchical,mantegna1999introduction,voit2005statistical,mantegna1999introduction,christoffersen2012elements,chakraborti2011econophysics,ferreira2017assessment}. To financial policy making,  manage risk, optimal portfolio, trading strategies, in addition to quantifying volatility, it is necessary to use robust statistical measures \citep{mantegna1999introduction,christoffersen2012elements,chakraborti2011econophysics,pagan1996econometrics,marti2021review}.   
	Quantifying the financial institutions from various aspects, which is a vital topic in economic development, has been examined in \citep{wen2016forecasting,gong2017forecasting,gong2018incremental,gong2018structural,he2019risk,yang2020systemic}.
	 
		There has been a great amount of research in analyzing the financial markets using the correlation, networks, clustering or models of measuring financial institutions’ tail risk, such as the conditional value-at-risk model. These methods are employed in analyzing the effect of geography on markets' return \cite{Drozdz}, the formation of clusters within a stock index
		\cite{mantegna}, and between geographically distant markets \cite{cross}. Complex Network Theory is  extensively carried out to find the key players of financial networks \cite{diebold2014network, yang2020systemic}. The construction of hierarchical structures between market sectors and indexes has also been investigated in \cite{Chakra, structure}. Analyzing the cross-correlation of return rates is performed in order to capture the markets' comovements and investigation of volatility spillover effects. Such analysis of cross-correlations over time is responsible for the evaluation of market dynamics and states \cite{Schafer, Asian, cool}, and it is broadly applied in understanding the market behavior during the crisis epochs \cite{Italo}. The core of this analysis is based on Random Matrix Theory, in which the eigenvalues of a correlation matrix constructed based on empirical data confronts the results obtained from a completely random matrix \cite{RMT1,RMT2}.
		Based on the Efficient Market Hypothesis, the variation of prices is a random process governed by a normal distribution; however, recent research suggests this hypothesis obsolete and claims that the variation of prices is statistically correlated \cite{Jin}. Subsequently, validating the deviation of markets from the Efficient Market Hypothesis is an important task. Choosing the proper statistics for analyzing the resulting time series, which can be assumed as Stochastic Fields, is a tricky task since each statistical framework has its advantages and disadvantages.
	
	 The notion of clustering has long-standing in analyzing complex systems \citep[and references therein]{mantegna1999hierarchical,marti2021review,Schafer,liao2005clustering,wu2013hierarchical,lahmiri2016clustering,jiang2008cluster,lahmiri2017clustering}. Specifically, the art of clustering of financial series mainly concentrates on constructing the adjacency matrix, including the distance (correlation coefficients) between any pairs of data sets, and then implementing the well-known minimum spanning tree \cite{huang2009comparison}. The widely adopted method is based on minimum spanning tree whose algorithms have been developed from various points of view ranging from preprocessing, calculating correlations, converting to distance, linkage clustering approaches to chaining phenomena and evaluation criteria \cite{marti2021review,liao2005clustering}. Dynamical clustering has been carried out by different methods such as computing correlation coefficient on a rolling window and evaluating temporal properties of constructed networks from underlying time series \cite{zhao2018stock}. Incorporating the clustering approaches provides valuable quantitative tools which enable us to examine the connection between financial markets and to address how the indices of markets evolve during the crisis epochs and volatility shocks.  
	 
	  In a large portion of  previous methods, the linear correlation between financial markets has been used, while the nonlinear relationship between assets essentially needs to incorporate more complicated measures such as distances based on Granger causality \cite{billio2012econometric}, partial correlation \cite{kenett2010dominating}, nonlinear relationship between financial institutions \cite{su2011non,wang2014dynamics}, mutual information \cite{ferreira2017assessment,fiedor2014information,rocchi2017emerging,kenett2010dynamics,fiedor2014networks,baitinger2017interconnectedness,barbi2019nonlinear,goh2018inference,guo2018development}, Copula-based \cite{marti2016optimal,durante2015cluster,brechmann2013hierarchical}, tail dependence \cite{lohre2020hierarchical}, (see also \cite[and references therein]{marti2021review}), leading to reliable results.

There are many methods ranging from topological and geometrical measures to examining probability density function and correlation function to characterize various data sets in $(1+1)$-, $(1+2)$-, and $(1+3)$-dimensions\footnote{Based on measure-theoretic approach, a ($n+D$)-dimensional random field (process) is a mapping from probability space to $\sigma$-algebra in which $n$ and $D$ are denoted to $n$-dependent variables and $D$-independent parameters, respectively \citep{adler81,adler2011topological,adler2010persistent}.}. Among various categories, excursion sets can be considered as significant features carried out not only for characterizing the morphology of underlying series but also for providing new approaches to extract characteristic scales from series.   A speculative definition for excursion sets can be written as: $\mathcal{A}_{\vartheta}(\mathcal{F})\equiv\{X|\mathcal{F}(X)\ge \vartheta \}$ \cite{adler81}. Where $\mathcal{F}(X)$ is a typical function of the underlying stochastic process and  $\vartheta$ is considered as a threshold. Such measures is capable of the magnification of deviation and discriminating the exotic features embedded in data sets.

A simple example of an excursion set is the so-called level crossing statistics introduced by S. O. Rice \cite{rice44a,rice44b}. Other modifications under the banner of Minkowski functionals are denoted by Up-, down- and conditional crossing statistics \cite{Bardeen:1985tr,Bond:1987ub,ryden1988,ryd89,mat96a,percy00,Matsubara:2003yt,tabar03,sadegh11,sadegh15,hadwiger2013vorlesungen}. Local extrema as a popular set of critical sets have been explored for Gaussian processes in $(1+D)$-dimension \cite{Bardeen:1985tr,vafaei2021clustering}. The one-point and two-point statistics of excursion sets provide complementary methods to quantify different series \cite{Bardeen:1985tr,matsubara,vafaei2021clustering}.	
The Two-Point Correlation Function (TPCF) of excursion and critical sets deal with how different values of the underlying systems occur relative to other values. The mentioned approach set up novel characteristic scales and extension of correlation values, particularly in non-linear form.   Pair correlation in terms of separation time (distance or angle) of a given feature is also known as clustering in some disciplines ranging from complex systems, condensed matter, granular materials, cosmology, biology, econophysics, to engineering \cite{young2001reproductive,peeb80,kaiser1984spatial,peac85,lumusden89,Bardeen:1985tr,Bond:1987ub,davis_peeb83,hamilton1993toward,szaouti98,hewet82,landy93,Marcos-Caballero:2015lxp,chakraborti2011econophysics}. 
 




 In this paper, we are going to look at clustering from a different perspective. More precisely, we will rely on the excursion features introduced to quantify the geometrical properties of typical stochastic processes. Based on the feature based approach, the cross-correlation coefficient will be computed and the hierarchical clustering method for the new measures will be applied. We will use our methods to 47 Stock market indices listed in Table \ref{index1} with the following advantages and novelties:
 	
 1) We will use the new features in the context of excursion and critical sets, namely local extrema and crossing measures. Comparison with the theoretical prediction for the Gaussian process to assess any deviation from Gaussianity or even exotic behavior will also be carried out.
 
 2) Going beyond the one-point statistics of geometrical measures and relying on the un-weighted two-point correlation function based on the excess probability of finding pairs of local extrema and crossing, we will accomplish the global and partitioning approaches. According to the clustering of excursion sets, the adjacency matrix for identifying states of financial market and temporal behavior of indices will be constructed. The structure of the matrix containing the excess probability of finding pairs of peaks is more capable of capturing occurrences of financial crises compared to the common correlation coefficient.

  The remaining of this paper is organized as follows: Section \ref{sec:math} gives the mathematical framework to set up the analytical prediction of peaks, troughs, and crossing number densities by using the spectral indices. Clustering of local extrema and up-crossing features according to the proper numerical estimators will be given in this section. The hierarchical clustering, more precisely, the agglomerative hierarchical clustering, will be explained briefly in section \ref{sec:math}. Data description is presented in section \ref{sec:data}. Section \ref{sec:results} is devoted to the implementation of the excursion set clustering on the financial markets in two approaches, namely globally and temporally. Section \ref{conclusion} gives discussions and conclusions.

	\section{Mathematical Framework}\label{sec:math}
	In this section, we introduce the most relevant techniques to analyze stock market data used as input series, based on excursion set theory.\\

	\subsection{Features clustering}
	
	The "clustering" notion is widely used in the context of stochastic fields \cite{lumsden1989clustering,rodriguez2014clustering}. Any mathematical criterion designed for revealing classification structures in a stochastic field represents a notion of clustering \cite{arabie1996clustering}.

		\begin{figure}
			\begin{center}
				\includegraphics[width=1\linewidth]{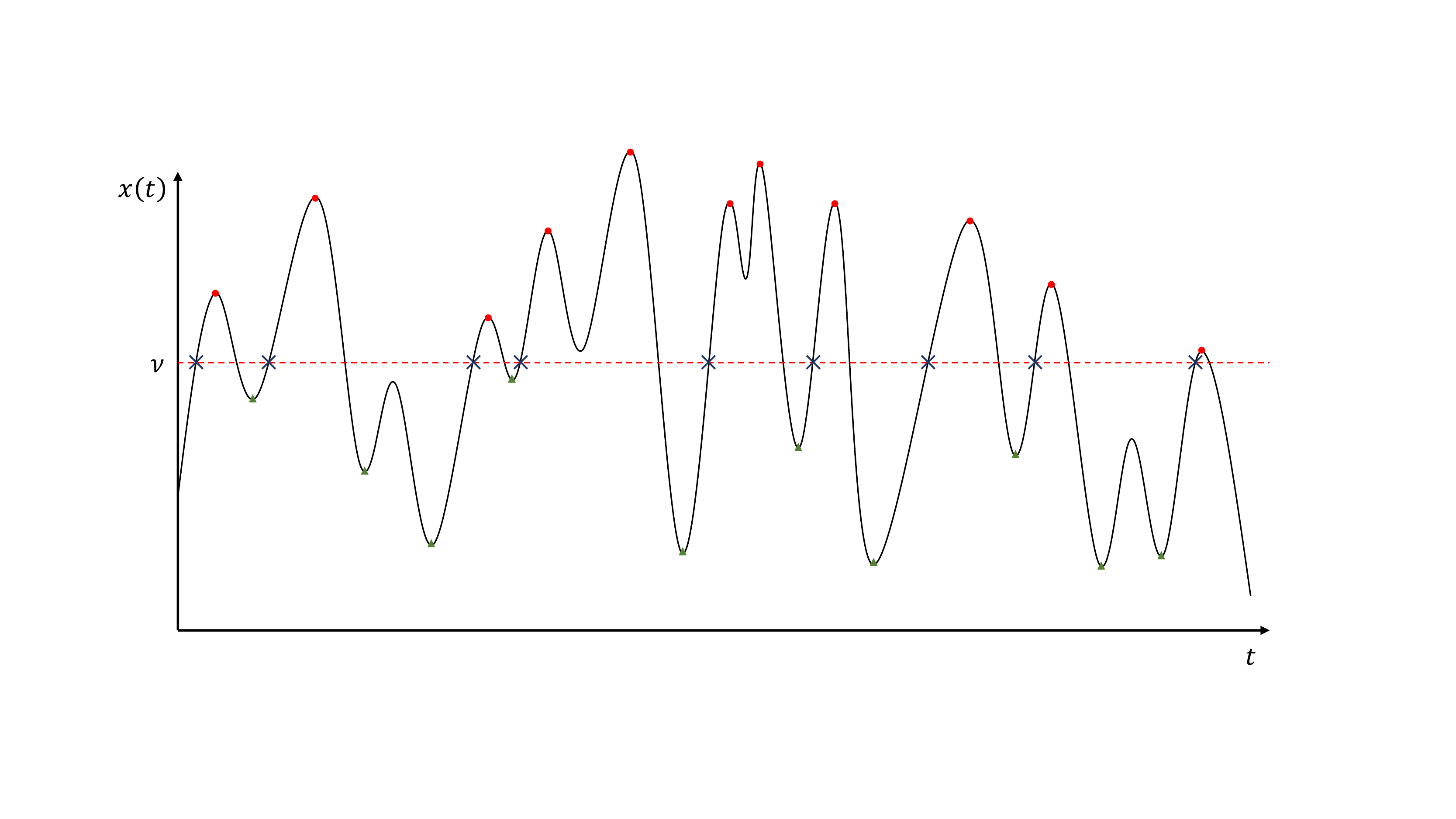}
				\caption{\label{up} The schematic representation of local maxima (filled circles), local minima (filled triangles), and up-crossing (cross) at a given threshold, $\vartheta$ for a smoothed time series.  }\label{fig:ptc}
			\end{center}
		\end{figure}

\begin{figure*}
		\begin{center}
			\includegraphics[width=0.5\linewidth]{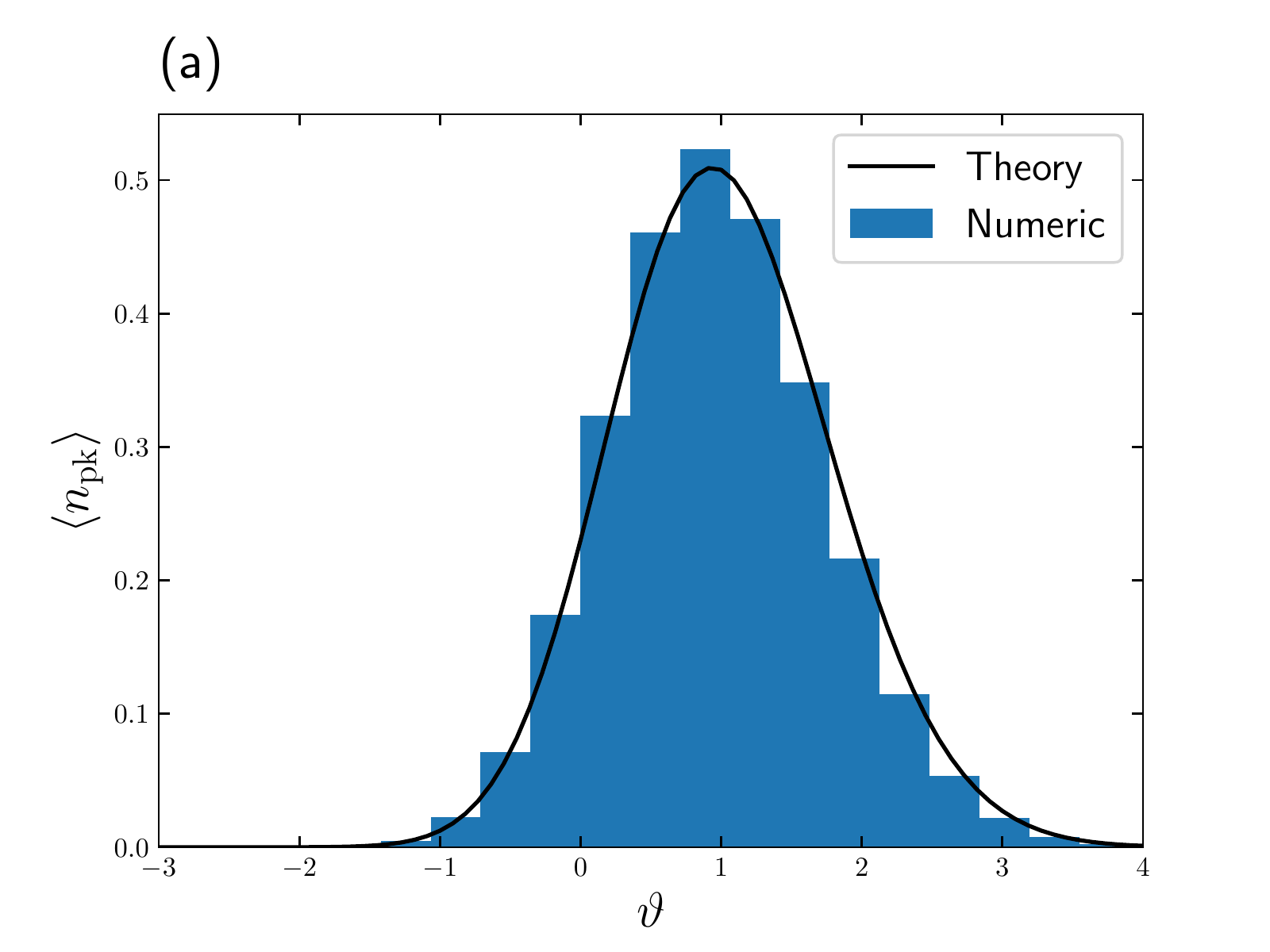}
			\includegraphics[width=0.5\linewidth]{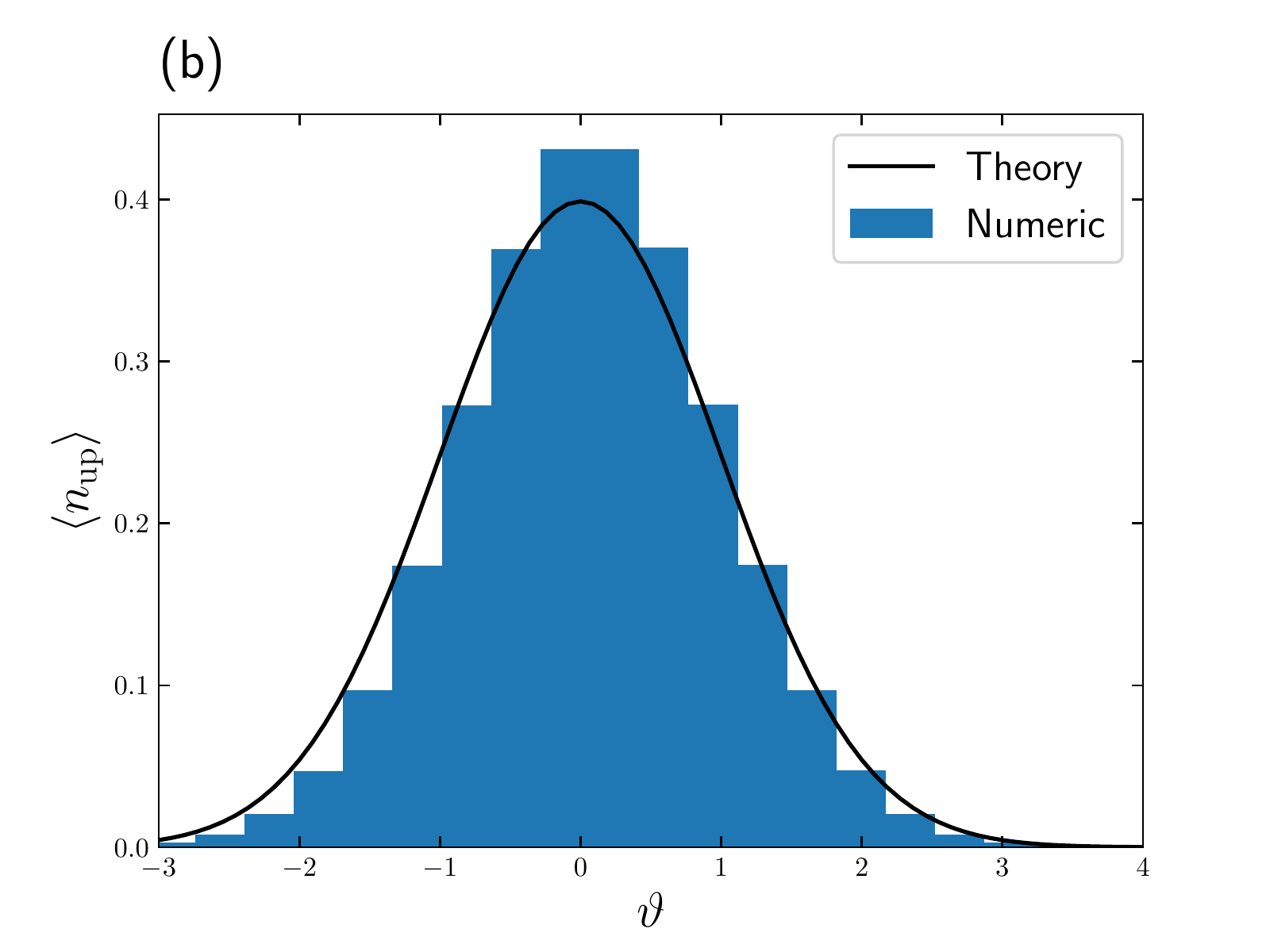}
			\caption{\label{fig:peakup} 
				Panel (a) and (b) correspond to the number density of peak and the number density of up-crossing as a function of $\vartheta$, respectively for a Gaussian white noise. The solid line is associated with the theoretical prediction.}
		\end{center}
	\end{figure*}

		There are many methods to figure out a feature in the underlying data set and then focus on examining corresponding properties \cite{tabar2019analysis}. A  robust method among them can be devoted to topological and geometrical measures to characterize statistical properties of a typical stochastic process, $\mathcal{F}$, in $n+D$ dimensions \cite{adler81,adler2011topological,adler2010persistent}. Here the index $n$ refers to $n$-dependent parameters, and $D$  is assigned to the $D$-independent parameters describing a $(n+D)$-Dimensional stochastic process. Local extrema (peak or trough) and crossing points are famous excursion sets in examining the morphological behavior \cite{rice44a,rice44b,peac85,lumusden89,Bardeen:1985tr,Bond:1987ub,sadegh15}. To clarify the mathematical definition of local extrema (peak and trough) as well as crossing points, we suppose that a continuous-time series\footnote{A well-defined approach to deal with a typical discrete time series is applying smoothing function in context of convolution to construct smoothed series \cite{matsubara}. } plays the role of our input data as $R(t)$. The number density of a desired feature  is given by: $n_{\diamond} \equiv\langle {\rm Conditions \; for \;  having \;  features} \rangle$, where $\diamond$ can be replaced by "pk" for peak (local maxima), "tr" for trough (local minima) and "up" for up-crossing (crossing with positive slope). To go further, we define a vector,  $\mathcal{A}$, containing all relevant quantities necessary for determining underlying features in our time series ($R(t)$). Throughout this paper, we consider  $\alpha(t)\equiv (R(t)-\langle R(t)\rangle)/\sigma_0$, $\eta\equiv(dR(t)/dt)/\sigma_1$ and $\zeta\equiv (d^2 R(t)/dt^2)/\sigma_2$  as normalized variables unless stated otherwise. The various order of spectral indices for a $(1+1)$-dimensional field read as: 
	\begin{equation}
	\sigma_n^2\equiv \frac{1}{2\pi}\int d\omega \omega^{2n}S(\omega)
	\end{equation}    
	where $S(\omega)$ is the power spectrum of $R(t)$. The mathematical description of local extrema number density in the threshold interval, $[\vartheta,\vartheta+d\vartheta]$ ($R=\vartheta\sigma_0$), becomes:   
	\begin{eqnarray}\label{eq:extrema1}
		\langle n_{\rm extrema}(\vartheta)\rangle=\frac{\sigma_2}{\sigma_1\sigma_0}\langle \delta_{\rm D}(\alpha(t)-\vartheta)\delta_{\rm D}(\eta)|\zeta|\rangle
	\end{eqnarray}
	here $\delta_{\rm D}$ is the Dirac delta function and $\langle \rangle\equiv \int {\rm d}^3 \mathcal{A}{\mathcal P}(\mathcal{A})$. The ${\mathcal P}(\mathcal{A})$ is the joint probability distribution of different components in $\mathcal{A}:\{ \alpha,\eta,\zeta\}$. 
	The domain of the second derivative for peak and trough are $\zeta \in (-\infty,0)$ and $\zeta \in (0,+\infty)$, respectively. Another interesting feature considered here is crossing statistics. Precisely, an up-crossing is defined by crossing with a positive slope when we move through time series at a given threshold. The mathematical description of up-crossing is:
	\begin{equation}\label{eq:up1}
		\langle n_{\rm up} (\vartheta)\rangle \equiv\frac{\sigma_1}{\sigma_0}\langle {\delta_D(\alpha(t)-\vartheta)\Theta(\eta)|\eta|} \rangle
	\end{equation}    
     	A schematic representation of local extrema and up-crossing of a continuous-time series is illustrated in Fig. \ref{fig:ptc}. For a Gaussian series in a stationary regime, the analytic form of peak (trough) number density in the threshold interval, $[\vartheta,\vartheta+d\vartheta]$, becomes \citep{Bardeen:1985tr,Bond:1987ub}:
	\begin{eqnarray}\label{eq:extrema22}	
	\langle n_{\rm pk}({\vartheta})\rangle=	\frac{\sigma_1\vartheta e^{-\frac{\vartheta^2}{2}}}{4 \pi
		\sigma _0^2} \left(1+ \text{erf}\left[\frac{\Gamma\vartheta}{\sqrt{2 (1-\Gamma^2)}}\right]\right)+ 
	\frac{\Gamma e^{-\frac{\vartheta^2}{2 (1-\Gamma^2)}}}{4 \pi ^{3/2}
		\sigma _1\gamma^2} \sqrt{2(1-\Gamma^2)}
	\end{eqnarray}	
where $\Gamma\equiv \frac{\sigma_1^2}{\sigma_0\sigma_2}$ and $\gamma\equiv\frac{\sigma_1}{\sigma_2}$ \citep{Bond:1987ub}.  
Also, the number density of up-crossing for a Gaussian time series in a stationary regime reads as: 
	\begin{equation}\label{eq:up2}
		\langle n_{\rm up}(\vartheta)\rangle=\frac{\sigma_1}{2\pi \sigma_0} e^{-{\vartheta}^2/2}
	\end{equation}    
	To achieve the number density above a threshold, the Dirac delta function is replaced by $\Theta(R(t)-\vartheta \sigma_0)$ ($\Theta$ is the Heaviside function). Fig. \ref{fig:peakup} illustrates the theoretical and numerical values of the number density of peak (panel a) and up-crossing (panel b) for a Gaussian white noise. The extension of number density of local extrema for weakly non-Gaussian processes can be found in \cite{matsubara2020statistics}.
	Our purpose in this paper is to go beyond one-point statistics. Accordingly, we can investigate more complementary statistical properties encoded in n-joint probability distribution functions. Two-point statistics, which is a trivial extension, is capable of quantifying the clustering notion of arbitrary features such as local extrema and crossing statistics. In the framework of the Two-Point Correlation Function (TPCF), two relevant measures can be considered to assess clustering:\\
	i) the weighted TPCF dealing with the autocorrelation; \\
	ii)  the unweighted TPCF, which estimates the excess probability of finding a pair of features by imposing proper conditions for a given time separation \cite{peeb80,kaiser1984spatial,peac85,lumusden89,Bardeen:1985tr,Bond:1987ub,davis_peeb83,hamilton1993toward,szaouti98,hewet82,landy93,Marcos-Caballero:2015lxp}.  The clustering of local extrema and crossing features separated by $\tau$ in a stationary time series at a threshold is written as: 
	\begin{equation}
		\xi_{\diamond-\diamond}(\tau;\vartheta)\equiv \frac{\langle n_{\diamond}(t+\tau; \vartheta)n_{\diamond}(t;\vartheta)\rangle}{\langle n_{\diamond}(t;\vartheta)\rangle\langle n_{\diamond}(t+\tau;\vartheta)\rangle}-1
	\end{equation}    
	in other words, the $\xi$ represents the excess probability of finding feature pairs, and $\xi+1$ corresponds to the pair correlation function. We expect that for a completely random process, $\xi_{\diamond-\diamond}(\tau;\vartheta)=0$. The mentioned expectation is also retrieved for general series for $\tau\to\infty$, where the correlation to be diminished. On the other hand, for small $\tau$, particularly for $\tau\to 0$,  one can not statistically find pairs of desired features separated by $\tau$, consequently $\xi_{\diamond-\diamond}(\tau\to 0;\vartheta)\to -1$. 
	Depending on the statistical properties of underlying data sets, the $\xi_{\diamond-\diamond}(\tau;\vartheta)$ illustrates different behavior for $0< \tau < \infty$,  causing discrimination between different processes using such clustering measure. In other words, $\xi_{\diamond-\diamond}(\tau;\vartheta)$ is the severity of accumulation and dissociation of "pk", "tr" and "up", compared to completely stochastic fields. The higher value of $\xi_{\diamond-\diamond}$ is essentially the higher value of clustering encoded in the data when we look for the existence of desired features (e.g., peaks, troughs, and up-crossing). The time scale associated with the maximum value of $\xi$ is so-called pair correlation characteristic time scale ($\tau_c$). Mentioned characteristics scale has a long history in examining an interacting system in the context of cluster expansion method \cite{mayer1940mayer,pathria}. Mentioned  time scale shows the statistical time interval for which the excess probability of finding feature pairs to be maximized, and it is useful to manage the crisis from the perspective of statistical physics.         

	The generalization of mentioned unweighted TPCF to incorporate two different features in one series or even two different data sets yields the unweighted two-point cross-correlation function (TPXF). This TPXF quantifies cross-correlation and provides more complete statistical descriptions.   
	
	Throughout this paper, we rely on the numerical computation of unweighted TPCF and TPXF of our data sets. To avoid the finite size effect and reduce probable spurious numerical results, some robust numerical estimators for determining $\xi_{\diamond-\diamond}(\tau;\vartheta)$ are introduced as follows:
	
	\begin{eqnarray} \label{eq:pp-estimator1}
		\xi^N_{\diamond-\diamond}(\tau;\vartheta)= \left(\frac{D_{\diamond}(t;\vartheta)D_{\diamond}(t+\tau;\vartheta)}{R_{\diamond}(t;\vartheta)R_{\diamond}(t+\tau;\vartheta)} \right) \frac{N_R^{\diamond}(N_R^{\diamond}-1)}{N_D^{\diamond} (N_D^{\diamond} -1)} -1\nonumber\\
	\end{eqnarray}
	\begin{eqnarray} \label{eq:pp-estimator2}
		\xi^H_{\diamond-\diamond}(\tau;\vartheta)= \frac{R_{\diamond}(t;\vartheta)R_{\diamond}(t+\tau;\vartheta)D_{\diamond}(t;\vartheta)D_{\diamond}(t+\tau;\vartheta)}{\left[D_{\diamond}(t;\vartheta)R_{\diamond}(t+\tau;\vartheta)\right]^2} -1\nonumber\\
	\end{eqnarray}
	\begin{eqnarray} \label{eq:pp-estimator3}
	\xi^{LS}_{\diamond-\diamond}(\tau;\vartheta)&=&\left( \frac{D_{\diamond}(t;\vartheta)D_{\diamond}(t+\tau;\vartheta)}{R_{\diamond}(t;\vartheta)R_{\diamond}(t+\tau;\vartheta} \right) \frac{N_R^{\diamond} (N_R^{\diamond} -1)}{N_D^{\diamond} (N_D^{\diamond} -1)} \nonumber\\
		&& - \left( \frac{D_{\diamond}(t;\vartheta)R_{\diamond}(t+\tau;\vartheta)}{R_{\diamond}(t;\vartheta)R_{\diamond}(t+\tau;\vartheta)} \right) \frac{N_R^{\diamond} (N_R^{\diamond} -1)}{N_D^{\diamond} N_R^{\diamond}} +1\nonumber\\
	\end{eqnarray}
	The $\xi^N_{\diamond-\diamond}$ is called the "natural estimator" \cite{Landy:1993yu}, and the $\xi^H_{\diamond-\diamond}$ has been introduced by  \cite{hamilton1993toward}, while the $\xi^{LS}_{\diamond-\diamond}$ is defined by \cite{Landy:1993yu}. In the above equation, $D_{\diamond}(t;\vartheta)D_{\diamond}(t+\tau;\vartheta)$ and $R_{\diamond}(t;\vartheta)R_{\diamond}(t+\tau;\vartheta)$ show the number of desired feature pairs in the data and the corresponding random sets, respectively. In the above equations, $N_D^{\diamond}$ and $N_R^{\diamond}$ are respectively the total numbers of local extrema or up-crossing in the data and random sets. In this paper, the given results have been computed by $\xi^N_{\diamond-\diamond}$, however, we checked other estimators and found that corresponding results are consistent together. We will use the above estimator and compute the clustering of peaks above a given threshold, troughs below a threshold, and up-crossing at the threshold. 
	
	
	\subsection{Hierarchical Clustering}
	
	Hierarchical clustering and dendrograms are powerful mathematical and pictorial tools for arranging elements (observations) and visualizing the underlying information in a set of various measurements. Here, we will give a brief explanation of the hierarchical clustering applied to our constructed matrices, whose elements are computed by the clustering notion based on the unweighted TPXF. To assess mentioned matrices, we use an agglomerative hierarchical clustering procedure (AHC). AHC initially assigns the $n$ nodes to the $n$ distinct clusters and merges each of the two clusters that are closest to each other determined by $\left(C_{i}, C_{j}\right)=\underset{\left(C_{k}, C_{l}\right)}{\arg \min } d\left(C_{k}, C_{l}\right)$, and iterates this procedure of fusion until one cluster is left which contains all of the nodes. There are several ways to define $d$ leading to different implementations of the AHC method \cite{clusterreview}. Among various schemes, we consider  Ward's minimum variance method \cite{ward}. Taking into account other routines for initiating AHC can however influence the final dendrogram depending on the similarity measures and distance update formulas. 	The inter-cluster similarity between clusters $A$ and $B$ that Ward suggested is defined by \cite{Mullner}:
	\begin{equation}
		d(A,B)\equiv\sqrt{\frac{2|A||B|}{|A|+|B|}}\cdot\left\|\vec{c}_{A}-\vec{c}_{B}\right\|_{2}
	\end{equation}
	where $\vec{c}_{\diamond}$ indicates the  centroid of a typical cluster denoted by $\diamond$ \cite{Mullner}. The distance update formula which is denoted by the distance between cluster $K$ and the newly formed $I \cup J$ cluster is defined by:
	\begin{equation}
		\label{LW}
		d(I \cup J, K)=	\sqrt{\frac{\left(n_{I}+n_{K}\right) d(I, K)+\left(n_{J}+n_{K}\right) d(J, K)-n_{K} d(I, J)}{n_{I}+n_{J}+n_{K}}}
	\end{equation}
	which is a specification of the Lance-Williams formula proposed by Lance G.N and Williams W.T \cite{LW} for updating the cluster dissimilarities after each iteration and the formation of new clusters. In Eq. (\ref{LW}) $I$ and $J$ are two clusters merged into one new cluster, and $K$ is any other cluster. Also, $n_I$, $n_J$, and $n_K$ are the number of the elements composing $I$, $J$, and $K$ respectively.
	The result of this method is a dendrogram visualizing a bottom-up hierarchy of nodes (i.e., markets) representing the similarity of markets' behavior in terms of clustering of critical points.

	\begin{figure}
		\begin{center}
			\includegraphics[width=0.8\linewidth]{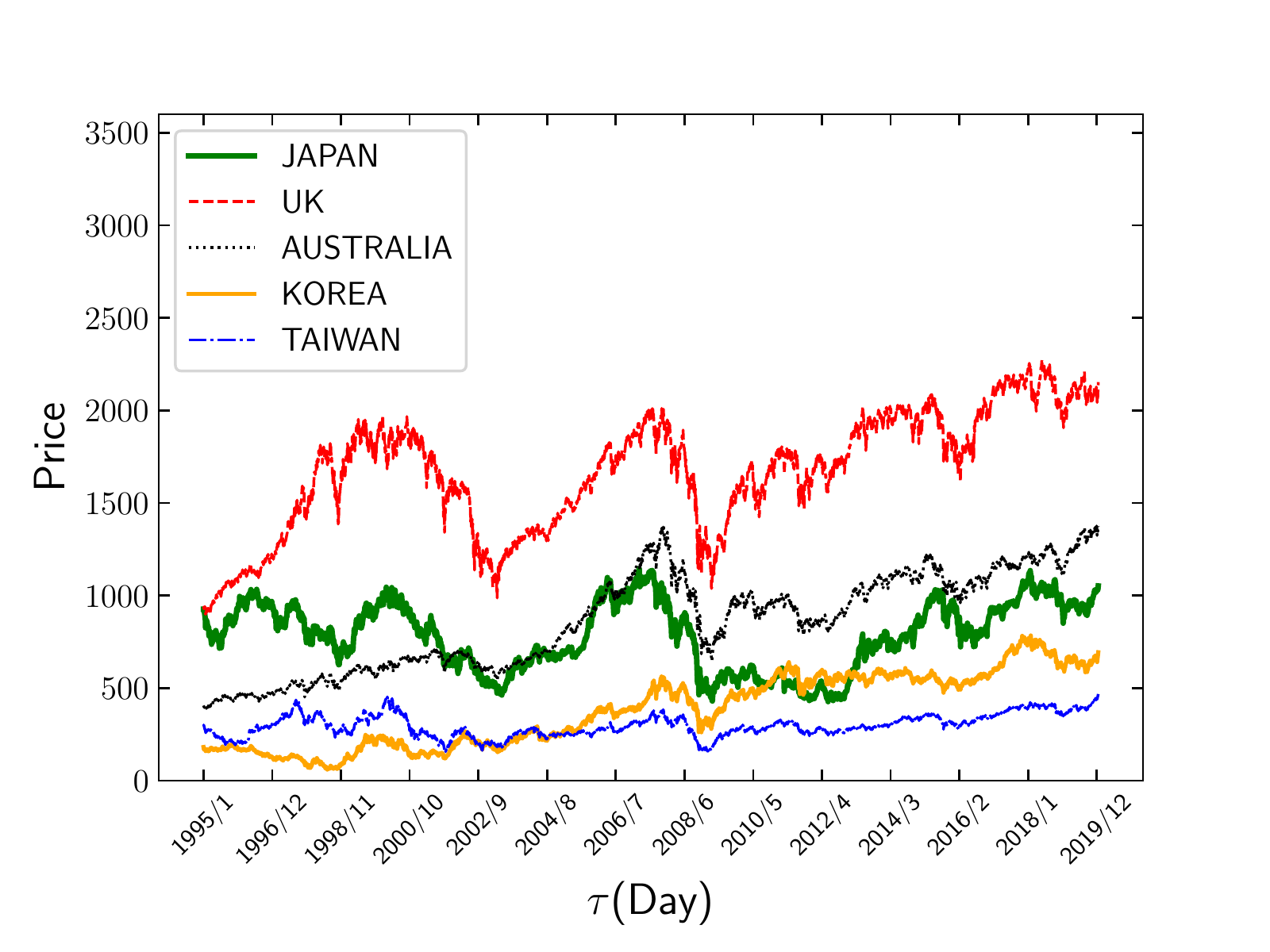}
			\caption{\label{fig:index} Time evolution of some daily stock market indices.}
		\end{center}
	\end{figure}
	
	\section{Data description}\label{sec:data}
	
	In this paper, we use the adjusted market capitalization stock market indices of 47 markets, classified into developed, emerging, and frontier markets. Mentioned data sets were constructed by Morgan Stanley Capital International (MSCI) and downloaded from DataStream (Eikon dataset). Our series per index are daily index prices over the period January 1995 to December 2019, corresponding to 6511
	observations \footnote{\texttt{http://www.msci.com}} (for more details on the different classification of our data sets see \cite{ferreira2017assessment}). To illustrate the general behavior of data sets, we depict the time evolution of some markets in Fig. \ref{fig:index}.

	\section{Implementation of Clustering on the Stock Indices}\label{sec:results}
	In this section, we apply the unweighted TPCF of local extrema and up-crossing features on the Stock data sets. To remove the well-known trend, we construct the log-return of series. We consider two approaches as follows:\\
	 i) The so-called global approach, in which, the time series for the whole time interval is considered. Our return data set is denoted by $R^{\ell}(k)$, where $\ell$ shows the label of the stock market running from 1 to 47 and $k$ represents the index of time.  \\
	 ii)  While in the second part, we divide each time series into non-overlapping segments containing 100 observed indices, known as the partitioning approach. For this part, our data sets is represented by $R^{\ell}_i(k)$, $i=1,...,65$ and $k=1,...,100$.    

	\subsection{Global approach}
	For the global approach, we calculate the one-point statistics and unweighted TPCF of local maxima and up-crossings for the whole time interval of our time series starting from January 1995 to December 2019. There have been several major incidents during this period, such as the 1997 East Asian Financial Crisis, 2008 Global Financial Crisis, etc., while the calculating $\xi_{\diamond-\diamond}(\tau, \vartheta)$ averages out many of these incidents. Nonetheless, since this data set consists of a very large number of markets with different economic structures, there is still valuable information in calculating these quantities, as they shed light on how the global economy behaves in a long run and how is the backbone of global financial markets.   

	\subsubsection{Number density of peaks and crossings}
\begin{figure}
			\begin{center}
				\includegraphics[width=0.55\textheight]{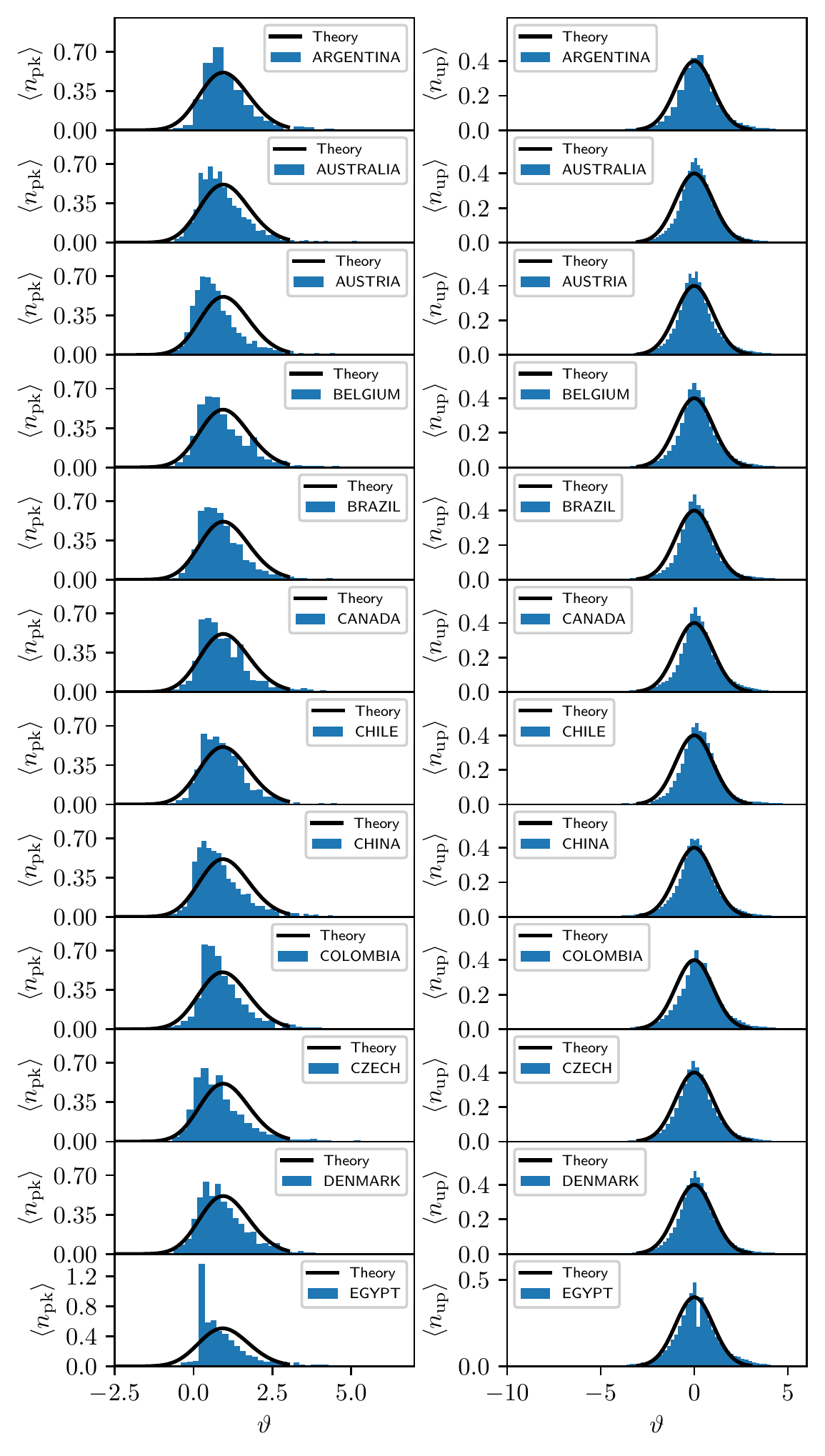}
				\caption{\label{fig:excursion1} The number density of peaks (left panel) and up-crossing (right panel) versus $\vartheta$ for Stock indices. The solid lines correspond to the theoretical prediction if the underlying data is Gaussian.}
			\end{center}
		\end{figure}

\begin{figure}
		\begin{center}
			\includegraphics[width=0.55\textheight]{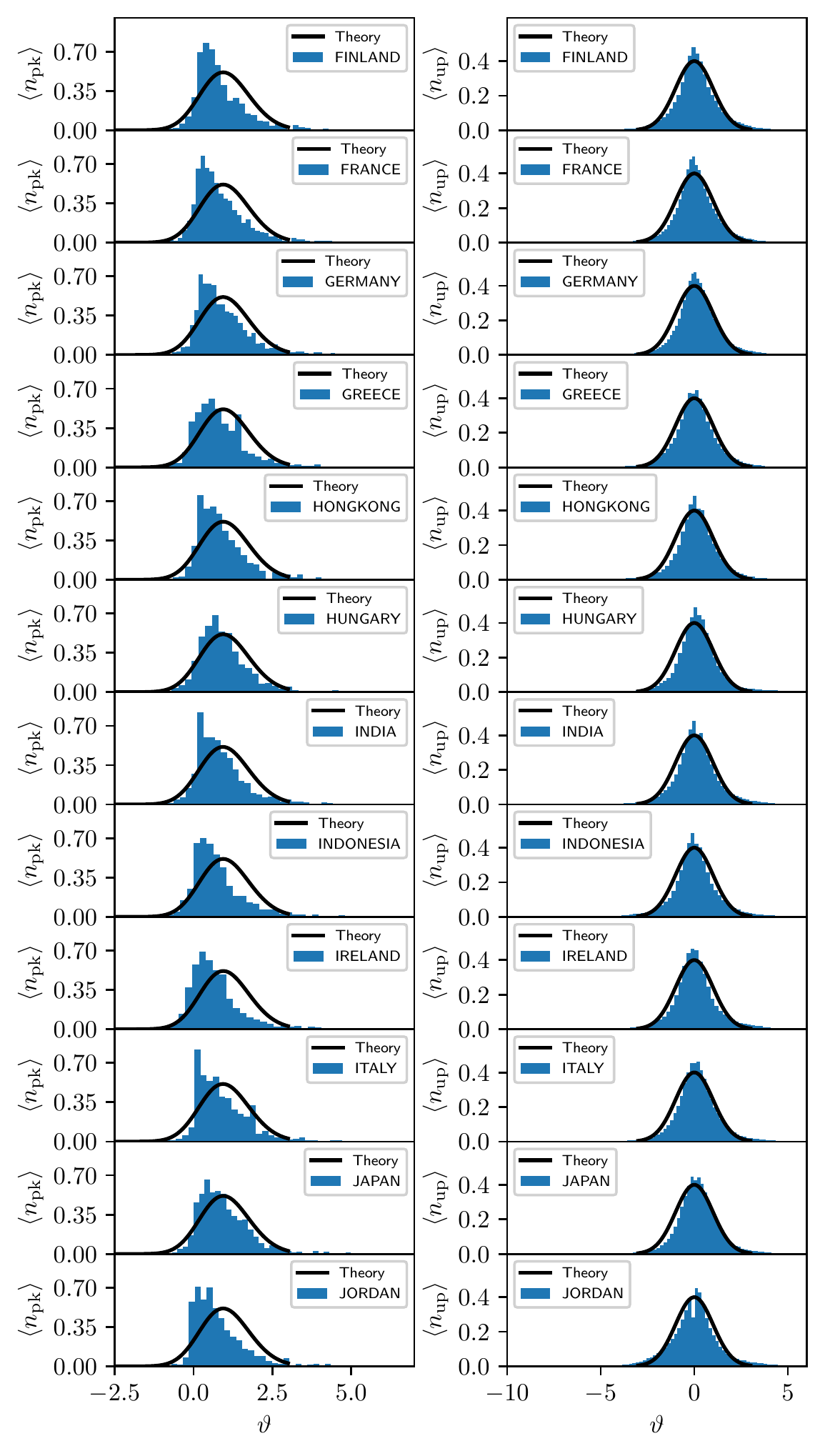}
			\caption{\label{fig:excursion2} Same as Fig. \ref{fig:excursion1}, just for other markets.}
		\end{center}
	\end{figure}

\begin{figure}
		\begin{center}
			\includegraphics[width=0.55\textheight]{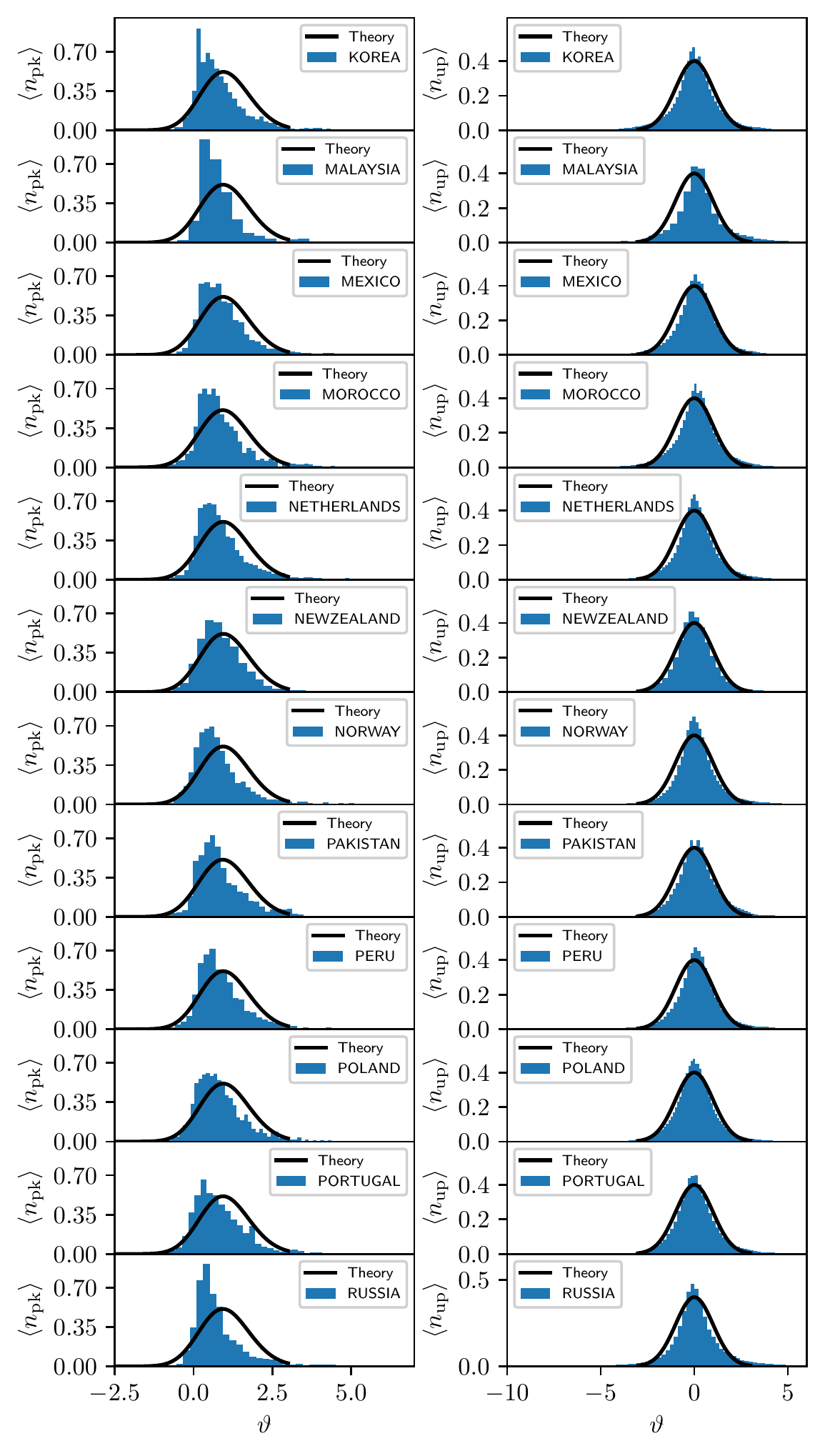}
			\caption{\label{fig:excursion3} Same as Fig. \ref{fig:excursion1}, just for other markets.}
		\end{center}
	\end{figure}

\begin{figure}
		\begin{center}
			\includegraphics[width=0.55\textheight]{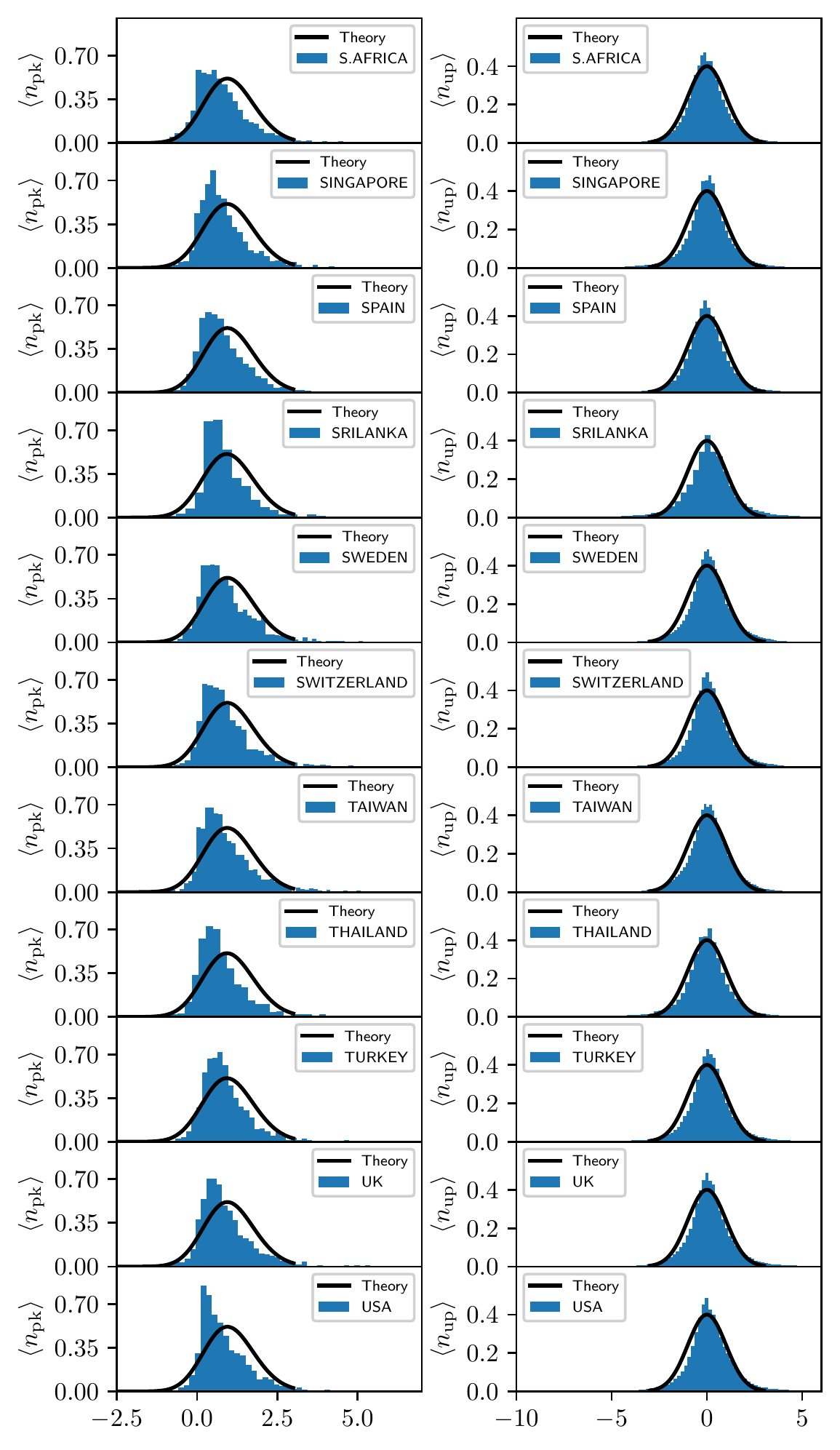}
			\caption{\label{fig:excursion4} Same as Fig. \ref{fig:excursion1}, just for other markets.}
		\end{center}
	\end{figure}

		Local maxima and crossing can be considered as the excursion sets. The former belongs to the critical point in a stochastic field and contains information regarding when and how a market's index rises and falls.  Before going further, we display the $\langle n_{\rm pk}\rangle $ and $\langle n_{\rm up} \rangle $ for some typical stock data sets in Figs \ref{fig:excursion1}-\ref{fig:excursion4}. Assessing the statistics of local extrema verifies that all markets studied in this paper for the given time interval have an excess value of local extrema density at a low threshold (almost around the mean value) with respect to the Gaussian theory. These results imply that the stock markets almost experience more ups and downs around mean value ($\vartheta\sim 0$) when they are evaluated for a long period. To highlight the meaning of up-crossing in its one-point statistics, we refer to research done by Jafari et. al., \cite{jafari2006level}. The authors argued that the inverse of the number density of crossing statistics is a characteristic time interval, within this time, the up-crossing will be observed again, statistically. In addition, the amount of up-crossing can be a measure to quantify the degree of development of the market \cite{jafari2006level}. The up-crossing number density represented in Figs. \ref{fig:excursion1}-\ref{fig:excursion4} demonstrate lake value at low threshold. It is worth mentioning that due to additional constraints accounted in local extrema compared to the up-crossing, the sensitivity of peak number density to capture the deviation from Gaussian theory is higher than up-crossing measure. Such deviations are observed for all markets and represents a universal property.      
			
	
		\subsubsection{Unweighted TPC(X)F of Local extrema and crossing}
\begin{figure*}
		\begin{center}
				\includegraphics[width=0.6\textwidth]{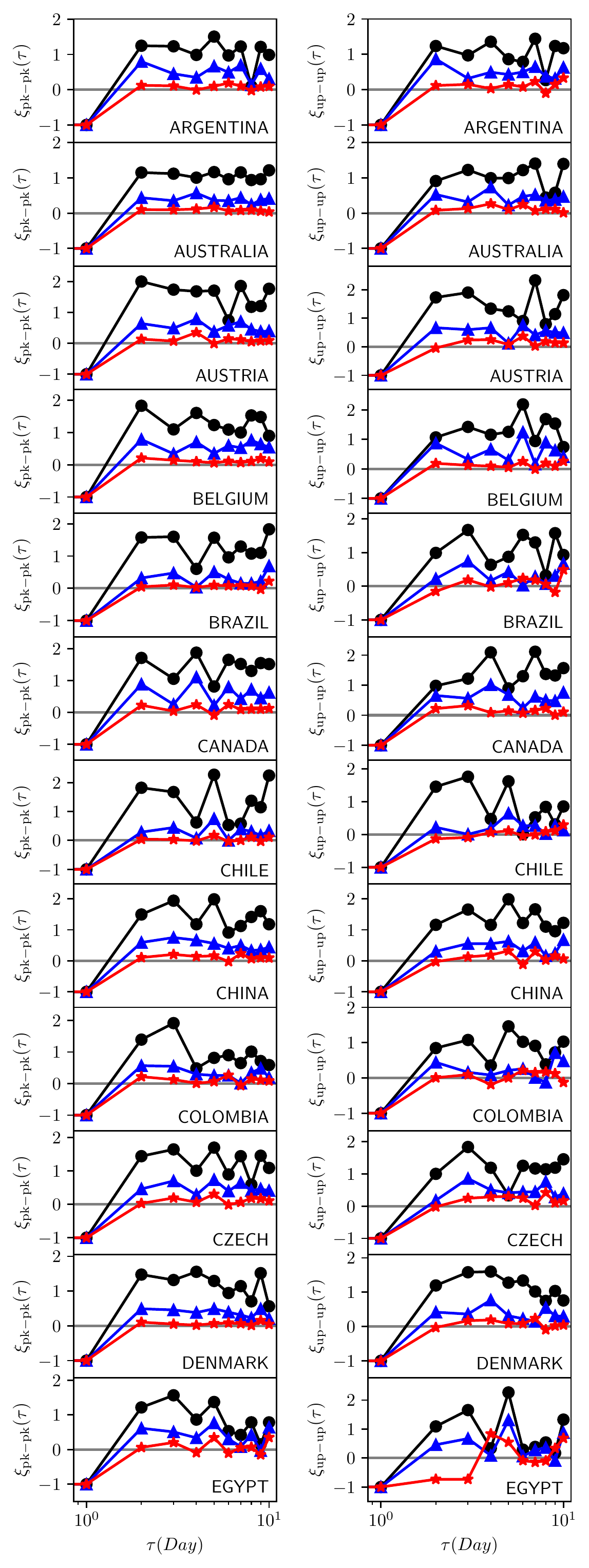}
	      	\caption{\label{DualTPCFPeaks1} The unweighted TPCF of the peak above a threshold (left panel) and up-crossing at the threshold (right panel) for stock indices. The filled circle corresponds to $\vartheta=+1\sigma$, the filled triangle shows the $\vartheta=0.5\sigma$ and the filled star is associated with $\vartheta=0\sigma$.}
		\end{center}
	\end{figure*}

\begin{figure*}
		\begin{center}
			\includegraphics[width=0.6\textwidth]{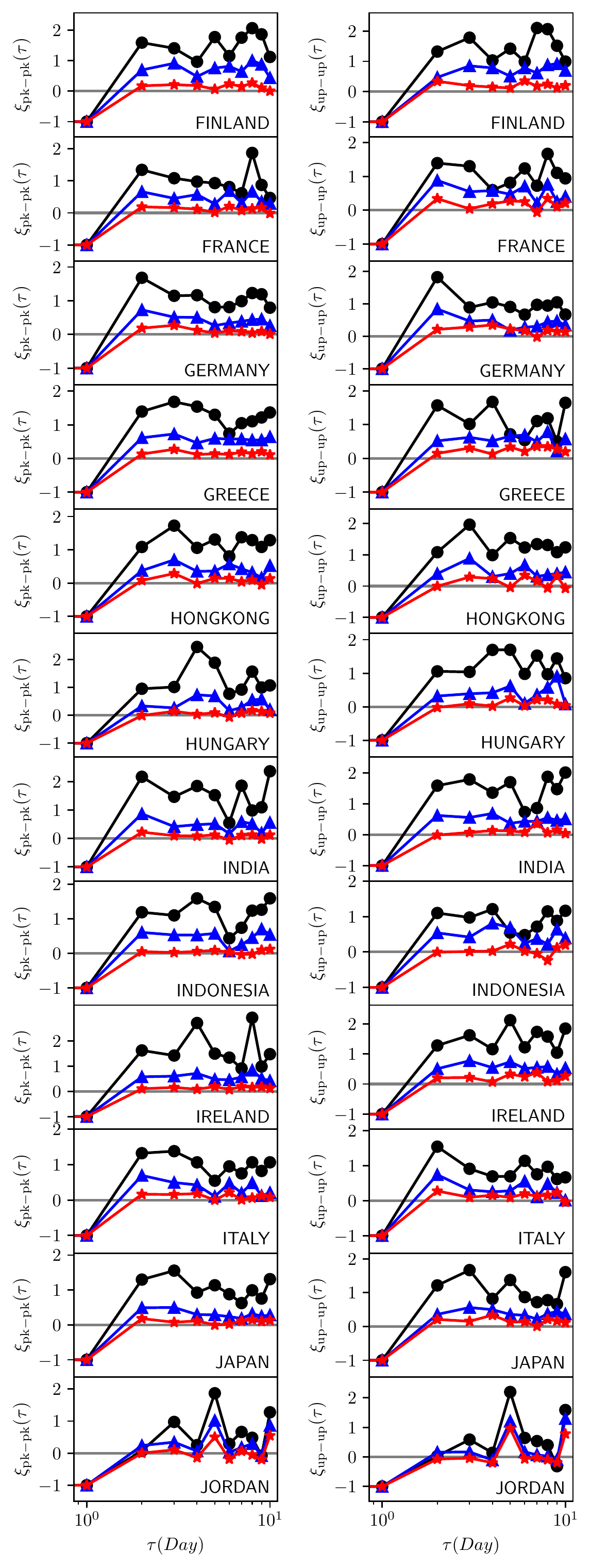}
			\caption{\label{DualTPCFPeaks2}Same as Fig. \ref{DualTPCFPeaks1}  just for other markets.}
		\end{center}
	\end{figure*}

\begin{figure*}
		\begin{center}
			\includegraphics[width=0.6\textwidth]{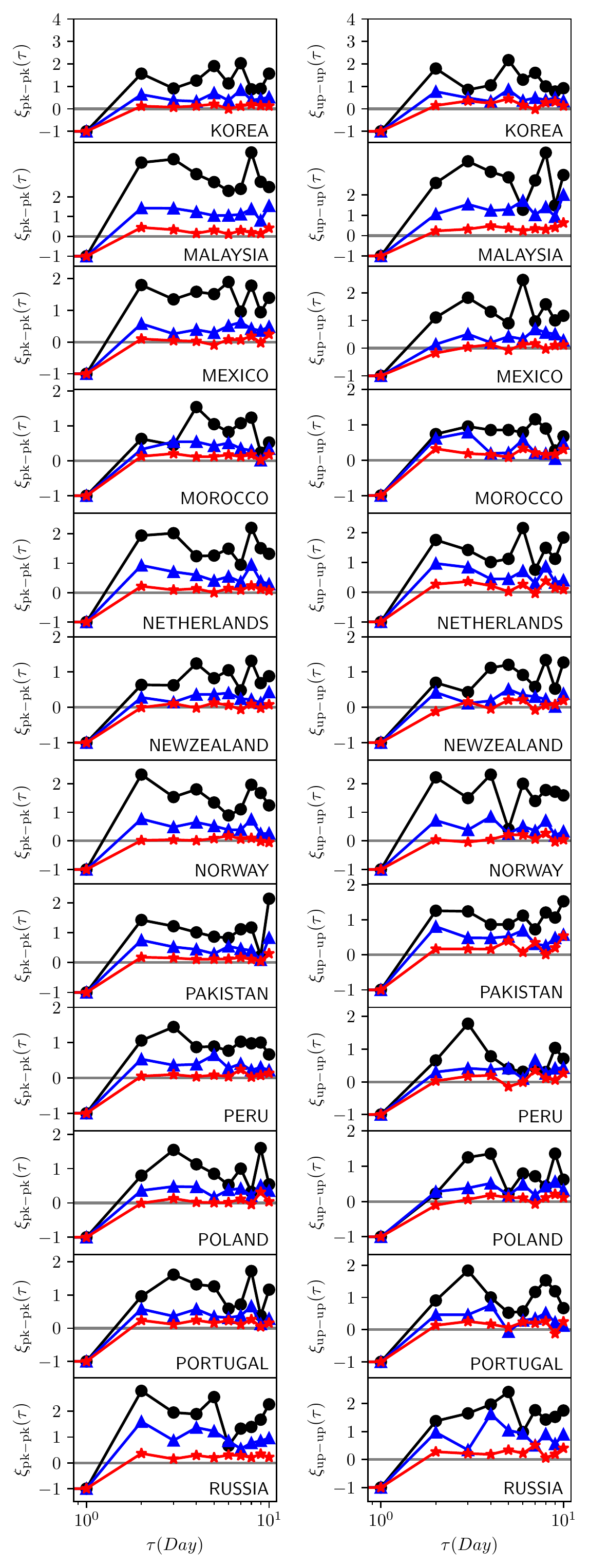}
			\caption{\label{DualTPCFPeaks3}Same as Fig. \ref{DualTPCFPeaks1}, just for other markets.}
		\end{center}
	\end{figure*}

\begin{figure*}
			\begin{center}
				\includegraphics[width=0.6\textwidth]{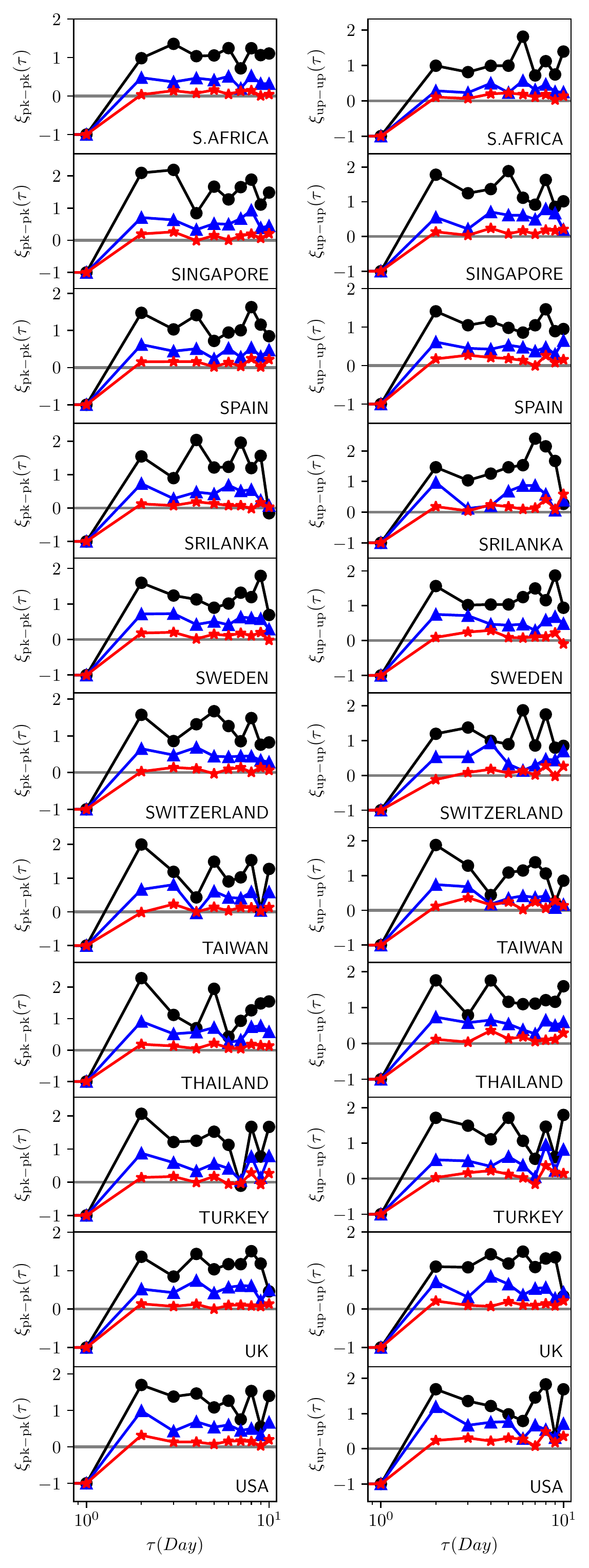}
				\caption{\label{DualTPCFPeaks4} Same as Fig. \ref{DualTPCFPeaks1}  just for other markets.}
			\end{center}
		\end{figure*}

\begin{figure*}
			\begin{center}
				\includegraphics[width=0.6\textwidth]{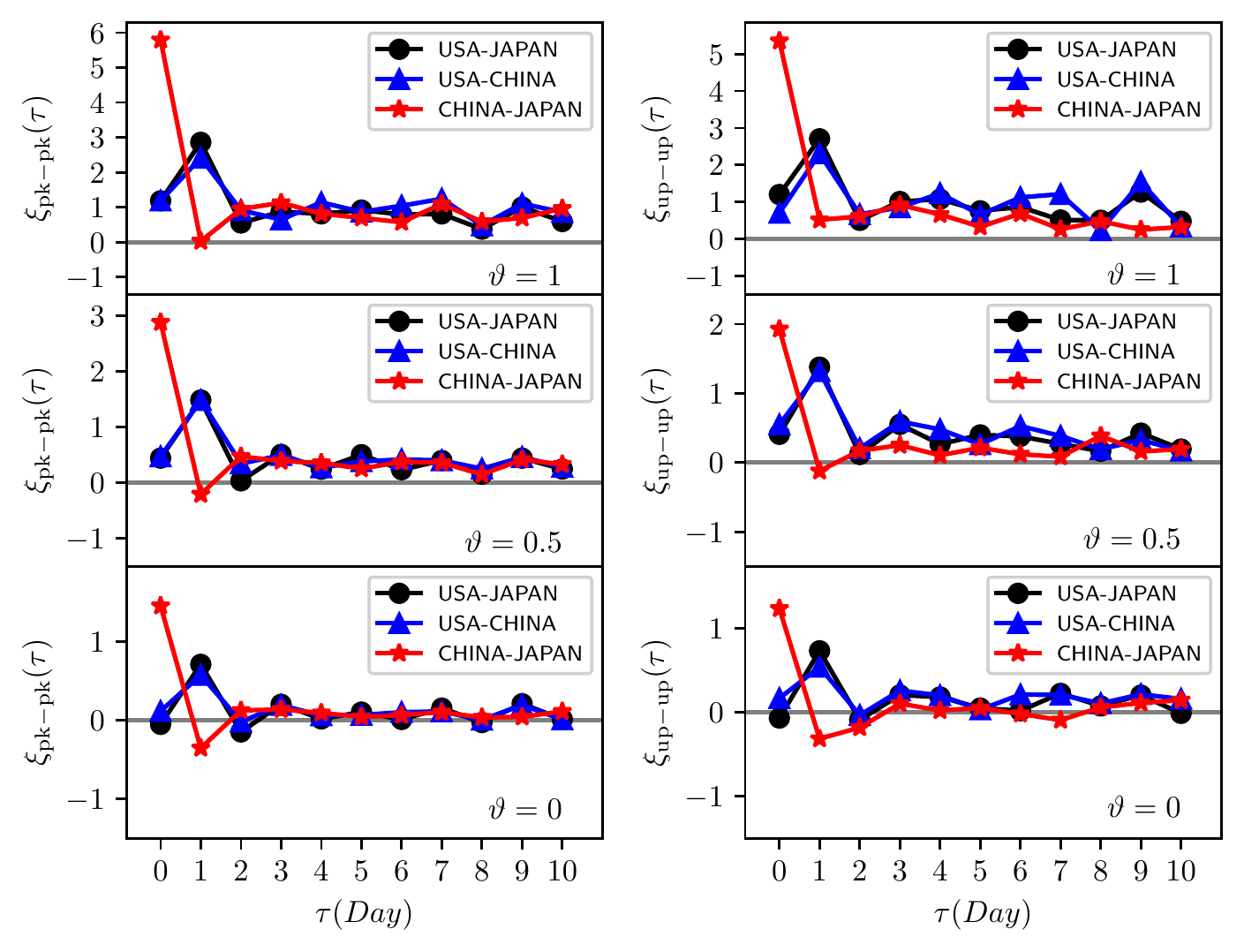}
				\caption{\label{crossTPCF} The unweighted TPCF of the peak above a threshold (left panel) and up-crossing at the threshold (right panel) for stock indices. This plot illustrates the behavior of cross-correlation between some typical indices.}
			\end{center}
		\end{figure*}

	Now we go beyond the one-point statistics and try to examine the clustering of excursion sets. To this end, we use the natural estimator (Eq. (\ref{eq:pp-estimator1})) to compute the unweighted TPCF of local maxima in our daily data set. Although, we have verified that considering other estimators indicated by Eqs. (\ref{eq:pp-estimator2}) and (\ref{eq:pp-estimator3}) yields consistent results. We also remove the mean value and transform the data to unit variance. To obtain statistically significant results, the peaks above a threshold have been considered throughout this paper. 
	Increasing the value of the threshold leads to having fewer points for computing unweighted TPCF, and the results have less statistical significance. The excess probability of finding a pair of local extrema happened with a given time separation for a completely random case essentially vanishes and irrespective of the statistical nature of the underlying series, we have $\xi_{\rm pk-pk}^{\ell \ell}(\tau\to 0)= -1$, which means that for small enough $\tau$, the probability of finding a pair of features for auto-correlation goes asymptotically to zero. At the intermediate regime, different behaviors emerge for various processes. In our case, by definition, the $\xi^{\ell \ell}(\tau=1{\rm Day})$ also equates to $-1$, because of data accusing made for daily recording. The unweighted TPXF of peaks, on the contrary, takes any values for a small enough time separation.    Figs. \ref{DualTPCFPeaks1} - \ref{DualTPCFPeaks4} illustrate the $\xi^{\ell \ell}_{\rm pk-pk}(\vartheta;\tau)$ (left column) and $\xi^{\ell \ell}_{\rm up-up}(\vartheta;\tau)$ (right column) for all data sets. In each plot, the filled circle, triangle and star symbols corresponds to $\vartheta=1\sigma_0$, $\vartheta=0.5\sigma_0$, and $\vartheta=0\sigma_0$, respectively. By increasing the threshold value, the excess probability of finding the pair of local extrema and crossing features grows, demonstrating a universal behavior and almost the maximum value of clustering statistically achieves during one week. Fig. \ref{crossTPCF} depicts the $\xi^{\ell \ell'}_{\rm pk-pk}(\vartheta;\tau)$ (left column) and $\xi^{\ell \ell'}_{\rm up-up}(\vartheta;\tau)$ (right column). Obviously, for markets located in the same geographical region, the associated unweighted TPXF achieves its maximum value for $\tau=0$, and for markets located in different geographical regions, we obtain a retarded impact on the behavior of excursion set clustering quantified by unweighted TPXF.

	Our results demonstrated that irrespective of the features considered for computing clustering in the stock markets, the unweighted TPC(X)F resembles almost the random behavior for $\tau\gtrsim 7$ days, therefore, the local extrema and crossing statistics contains relatively short term memory compared to the common weighted TPCF. For excursion sets, besides the value of the underlying series, some additional proper conditions should be satisfied. Consequently, the joint PDF of the value of series accompanying conditions vanishes faster than the marginalized joint PDF over mentioned conditions.

\begin{figure}
		\begin{center}
			\includegraphics[width=0.50\textwidth]{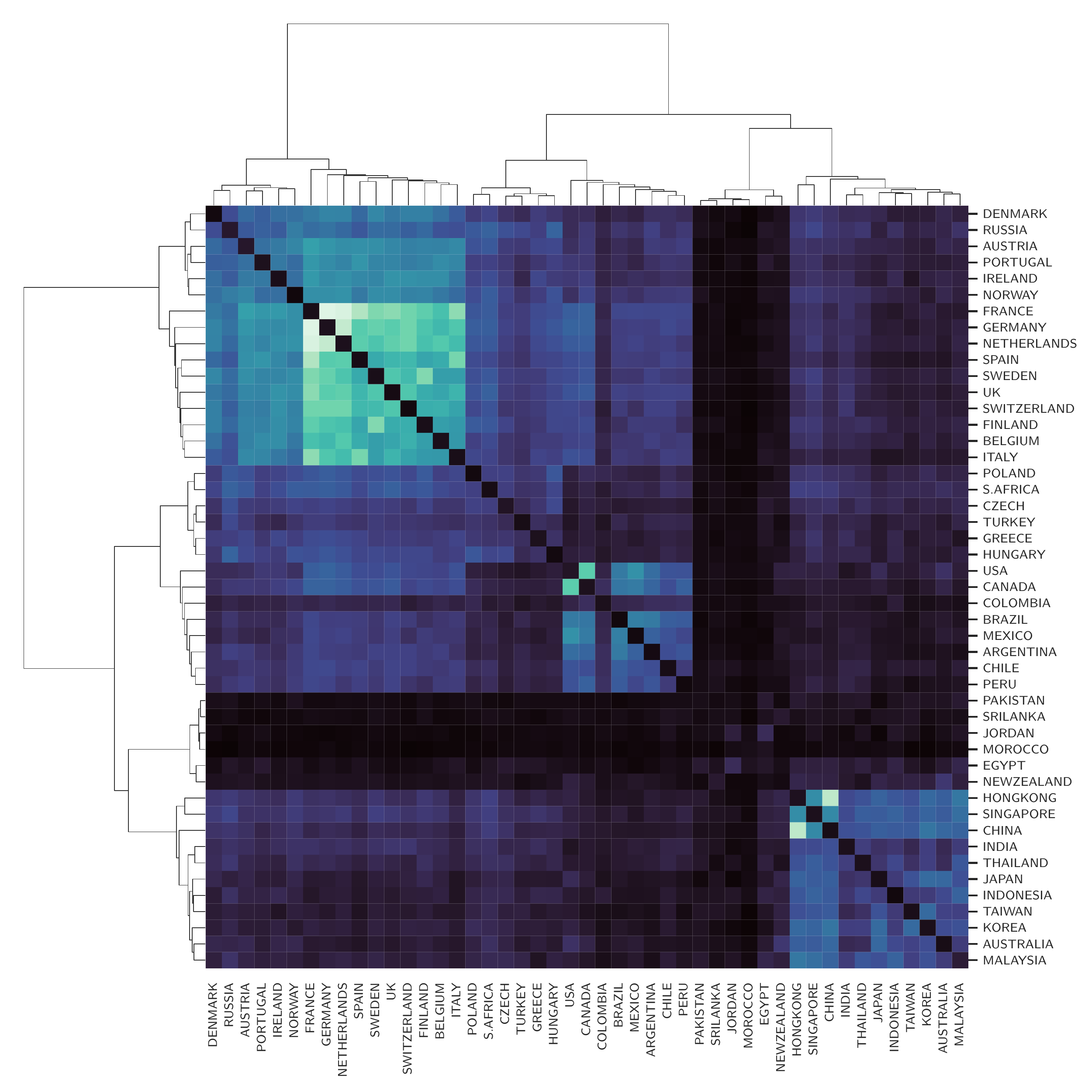}
			\includegraphics[width=0.50\textwidth]{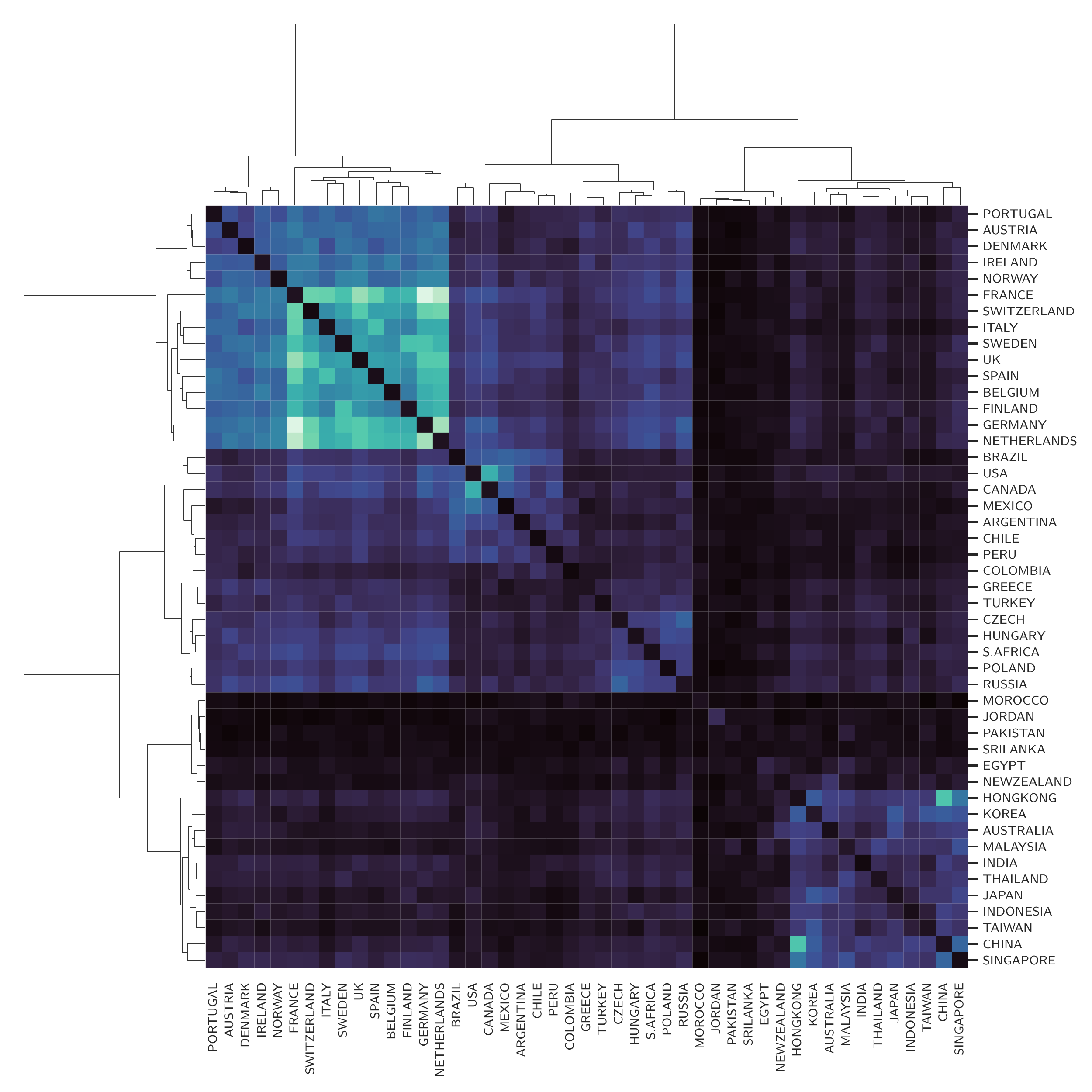}
			\includegraphics[width=0.50\textwidth]{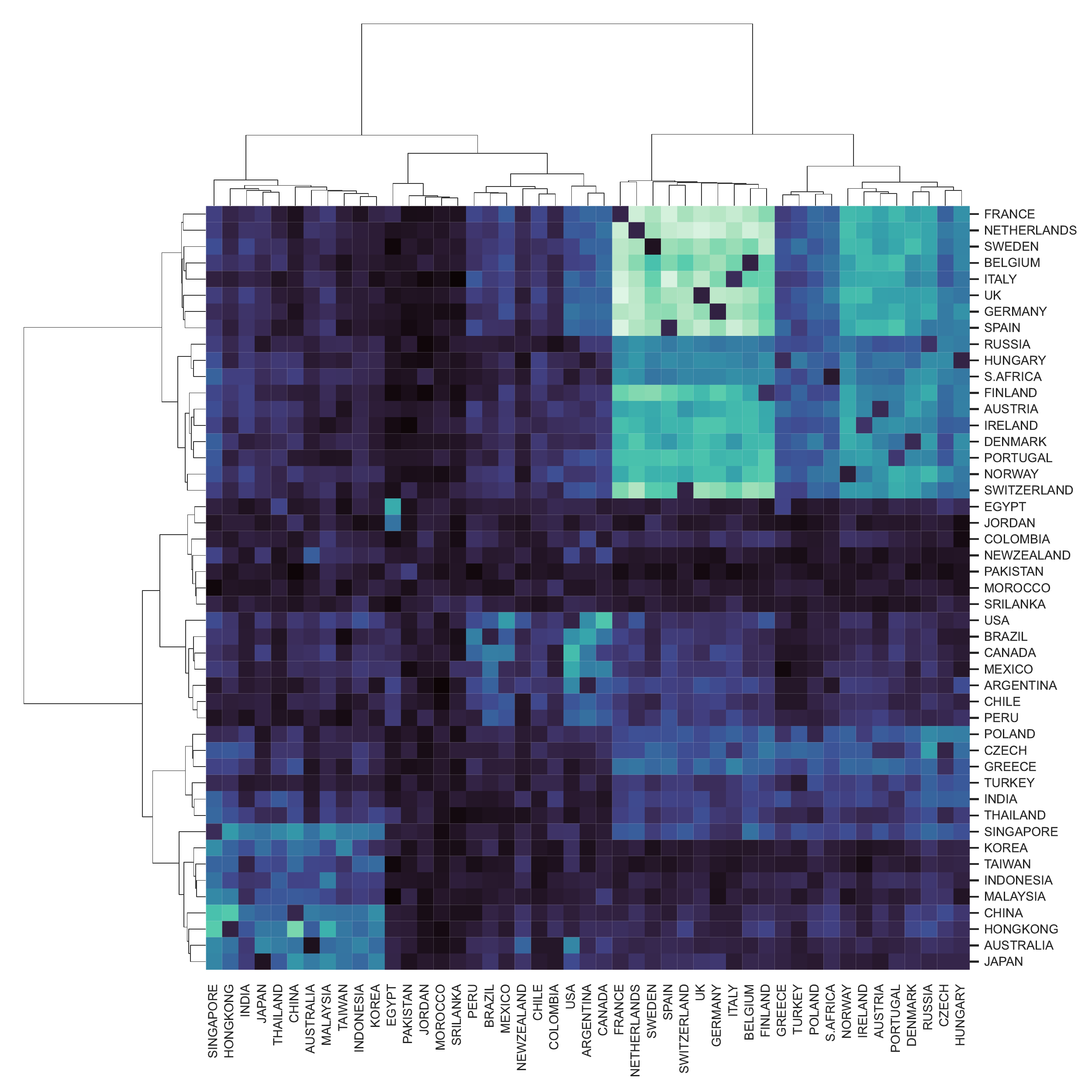}
			\caption{\label{fig:AHC1} AHC implementation on the $\Psi^{\rm pk}_{\rm max}\equiv\xi^{\rm max}_{\rm pk-pk}(\vartheta\ge 0\sigma_0)+1$ (upper panel), $\Psi^{\rm up}_{\rm max}\equiv\xi^{\rm max}_{\rm up-up}(\vartheta= 0\sigma_0)+1$ (middle panel), and $\Psi^{\rm pk-tr}_{\rm max}\equiv\xi^{\rm max}_{\rm pk-tr}(\vartheta_{\rm pk}\ge 0\sigma_0;\vartheta_{\rm tr}\le 0\sigma_0)+1$ (lower panel).   }
				\end{center}
	\end{figure}

 To make our analysis more complete in terms of cross-correlation, we also compute $\xi^{\ell \ell'}_{\rm pk-pk}(\tau, \vartheta)$ for $i$th and $j$th Stock markets, where $\ell,\ell'=1,...,47$. Accordingly, we construct a matrix denoted by $\Psi^{\rm pk}_{\ell \ell'}(\tau;\vartheta)\equiv \xi^{\ell \ell'}_{\rm pk-pk}(\tau;\vartheta)+1$ with size $47\times 47$. Now, we turn to the agglomerative hierarchical clustering procedure to identify the state of the stock market when the unweighted TPXF of the underlying series is taken into account. We construct the $\Psi^{\rm pk}_{\rm max}$ matrix whose elements are the maximum value of $\xi_{\rm pk-pk}(\vartheta\ge 0\sigma_0;\tau)$. The upper panel of Fig. \ref{fig:AHC1} shows the AHC results for $\Psi^{\rm pk}_{\rm max}(\vartheta\ge 0\sigma_0)$. The footprint of the geographical region in making more coherent the stock indices is obviously recognized in this approach. The middle panel of Fig. \ref{fig:AHC1} illustrates the same analysis just for $\Psi^{\rm up}_{\rm max}(\vartheta= 0\sigma_0)$. This geometric feature behaves the same as local maxima. The lower panel of Fig. \ref{fig:AHC1} also displays the cross-analysis of features, namely $\Psi^{\rm pk-tr}_{\rm max}(\vartheta_{\rm pk}\ge 0\sigma_0;\vartheta_{\rm tr}\le 0\sigma_0)$. This kind of measure also can classify the stocks in some categories in which the impact of the geographical region has considerable influence on the clustering of stock markets. However, looking at Fig. \ref{fig:AHC1}, more precisely, we obtain that the peak statistics is more capable of capturing blocks distinctively in the matrix identifying different groups. However, various measures, namely excursion and critical sets, find market structures compatibly. In the cross-analysis, the computed matrix is not symmetric by definition, nevertheless, the clustered stocks are almost in agreement with the results given by the Peak-Peak method. The imprint of threshold value has been investigated, and almost the same results have been obtained.  
 
\begin{figure*}
		\begin{center}
			\includegraphics[width=0.24\textwidth]{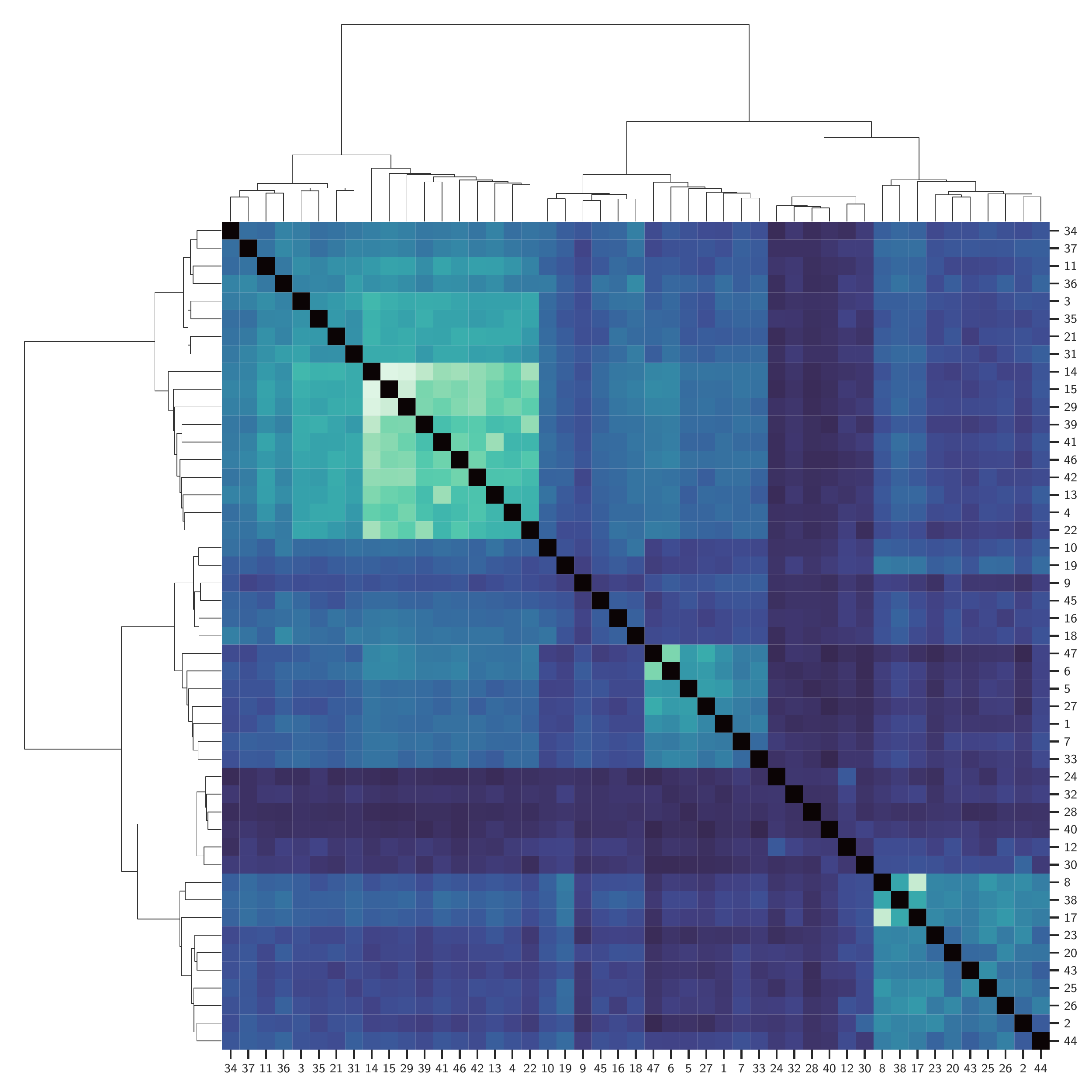}
			\includegraphics[width=0.24\textwidth]{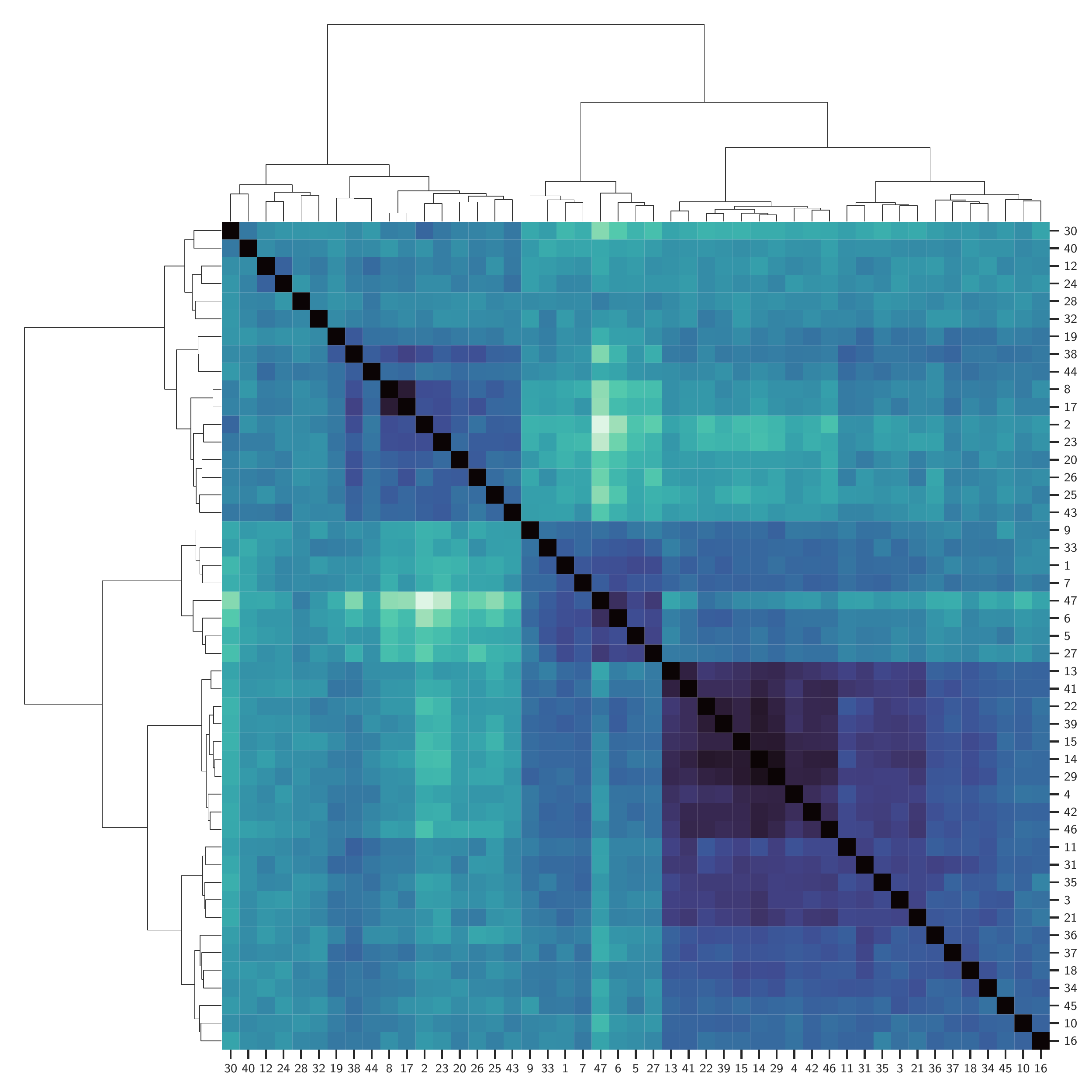}
			\includegraphics[width=0.24\textwidth]{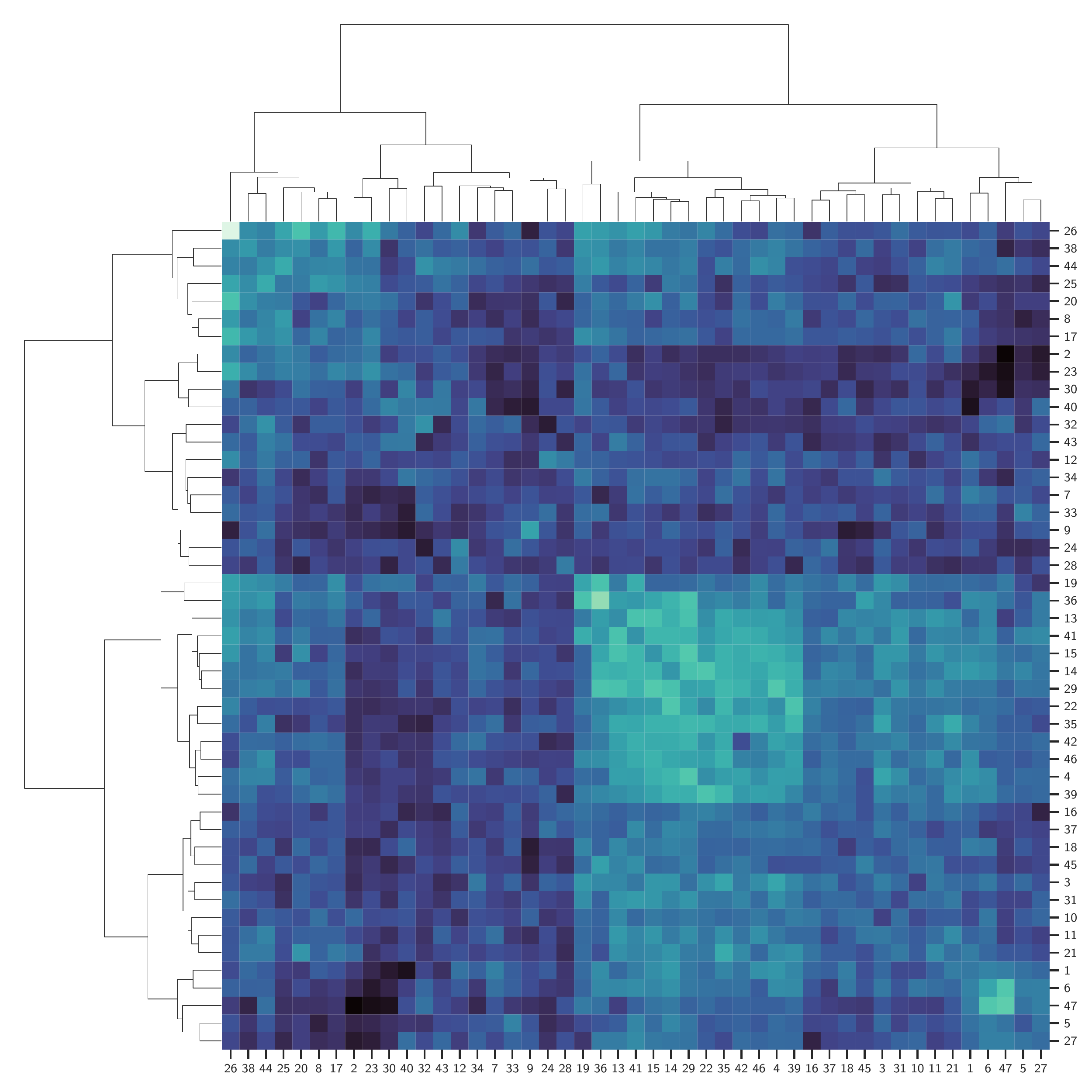}
			\includegraphics[width=0.24\textwidth]{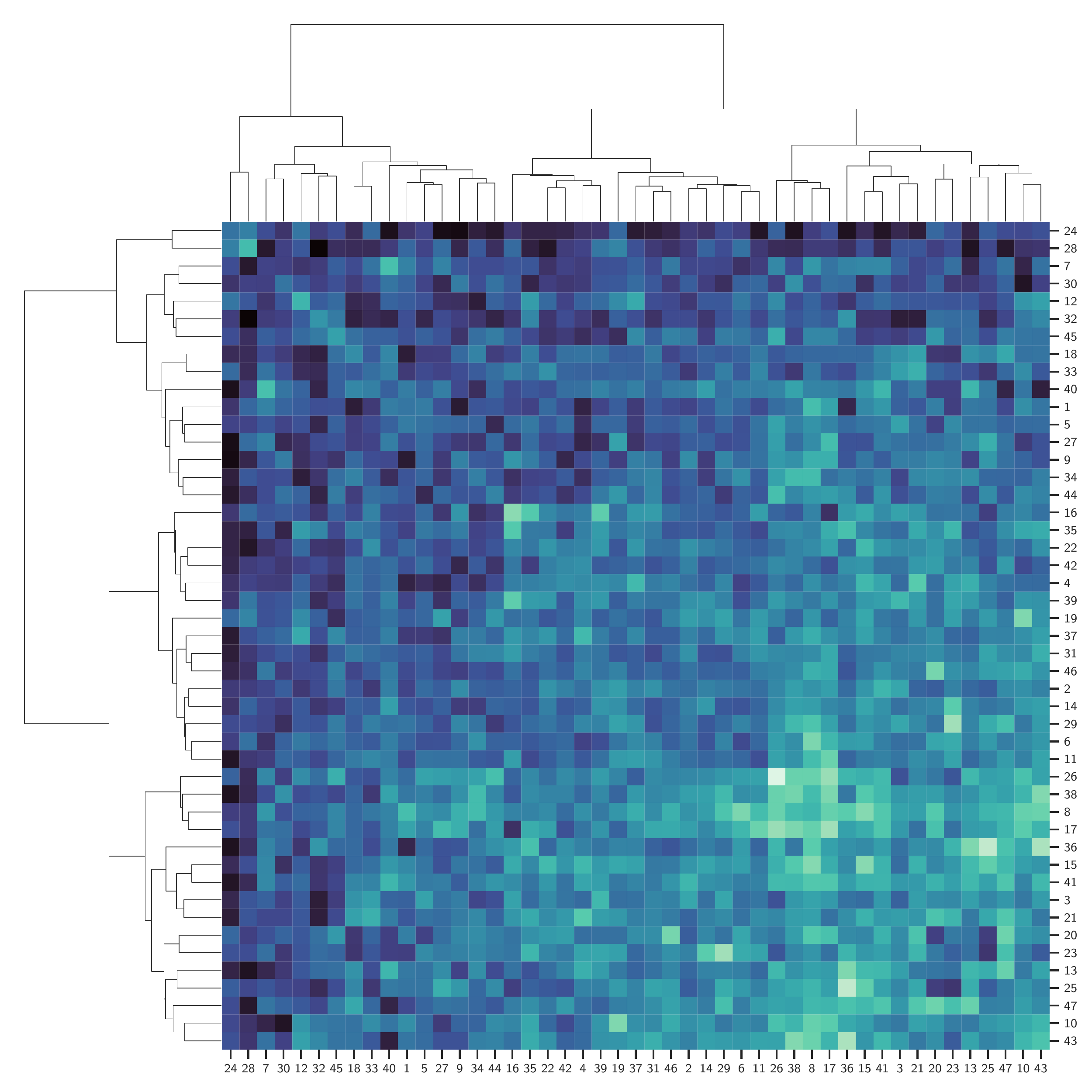}
			
		    \includegraphics[width=0.24\textwidth]{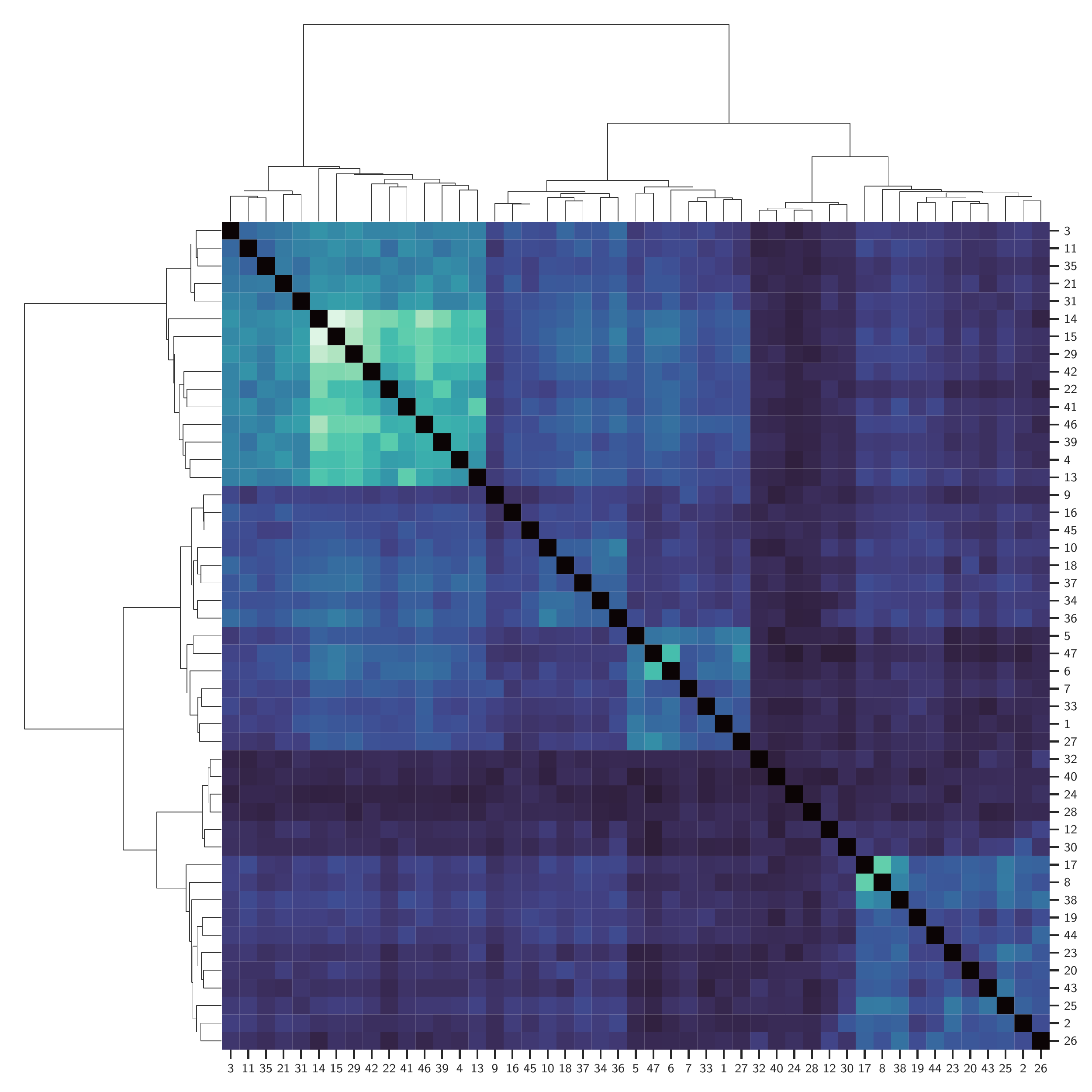}
		    	\includegraphics[width=0.24\textwidth]{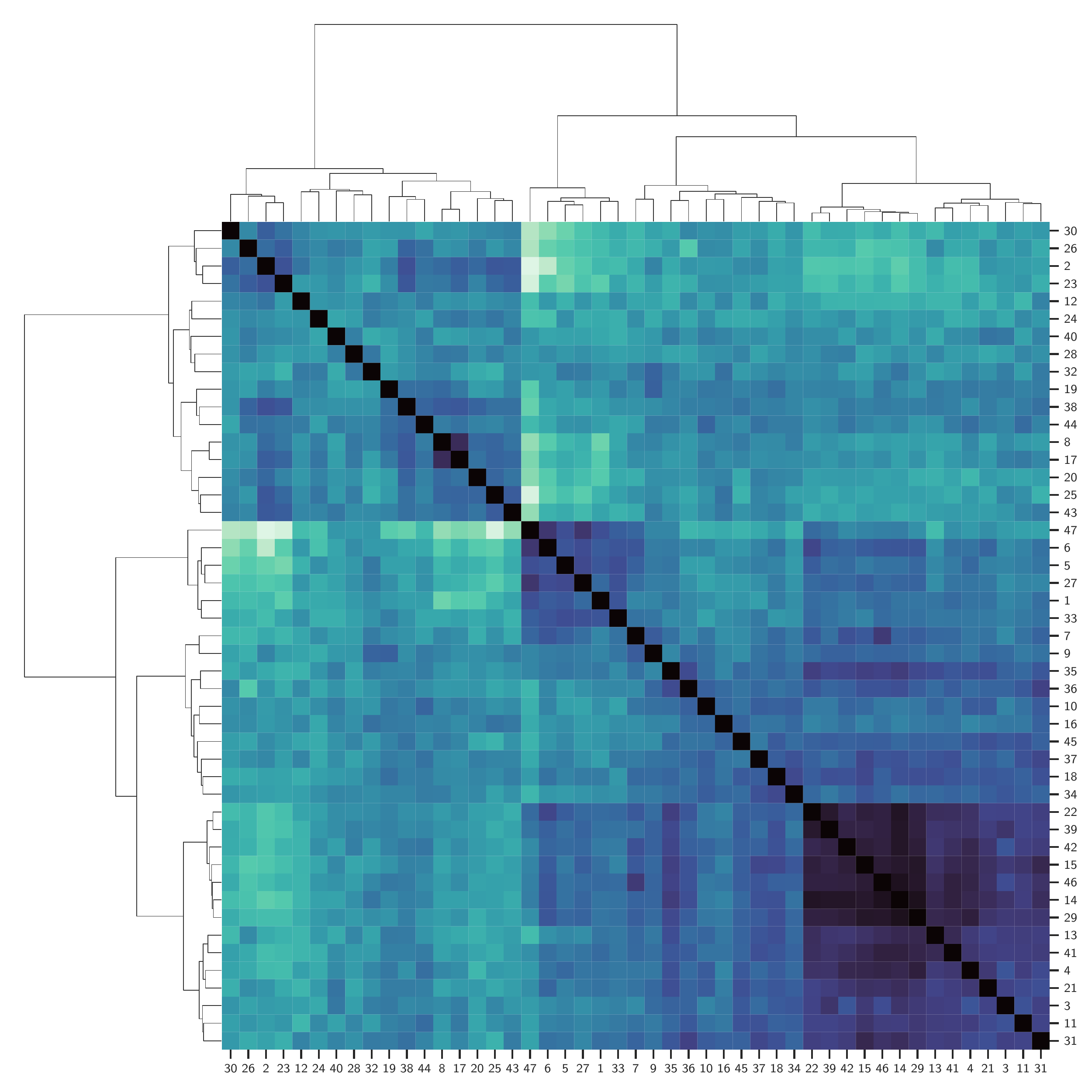}
		    		\includegraphics[width=0.24\textwidth]{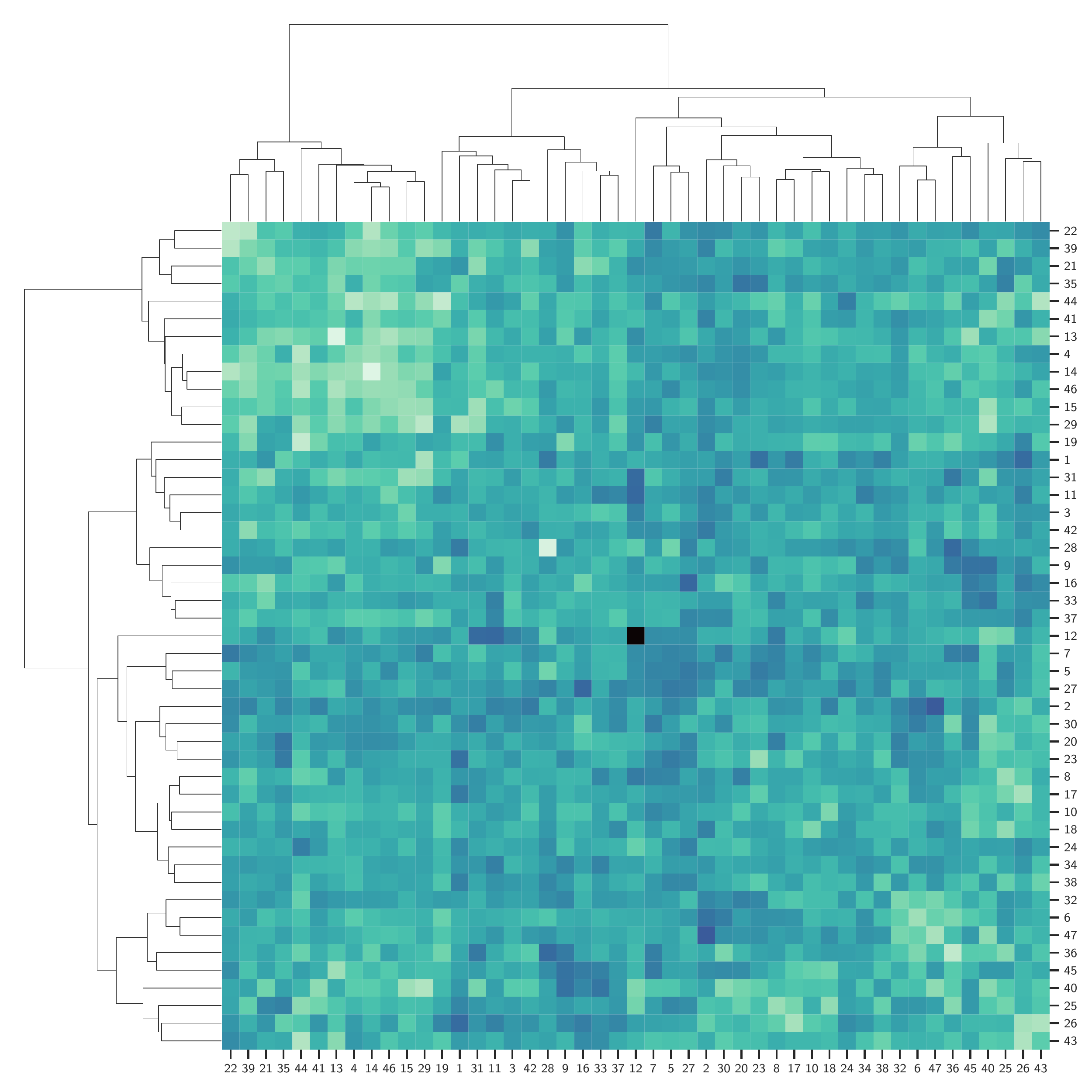}
		    			\includegraphics[width=0.24\textwidth]{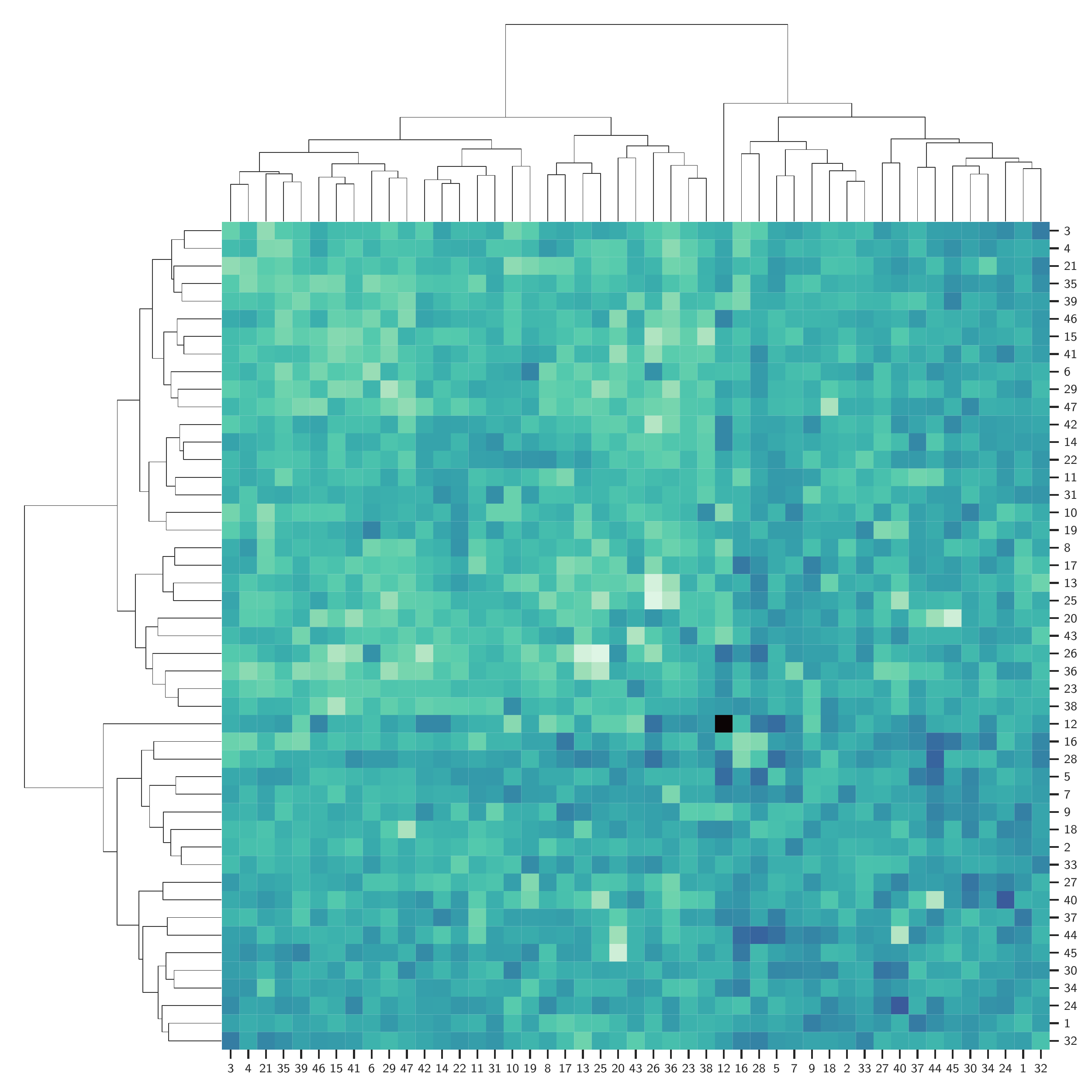}
		    			
					\includegraphics[width=0.24\textwidth]{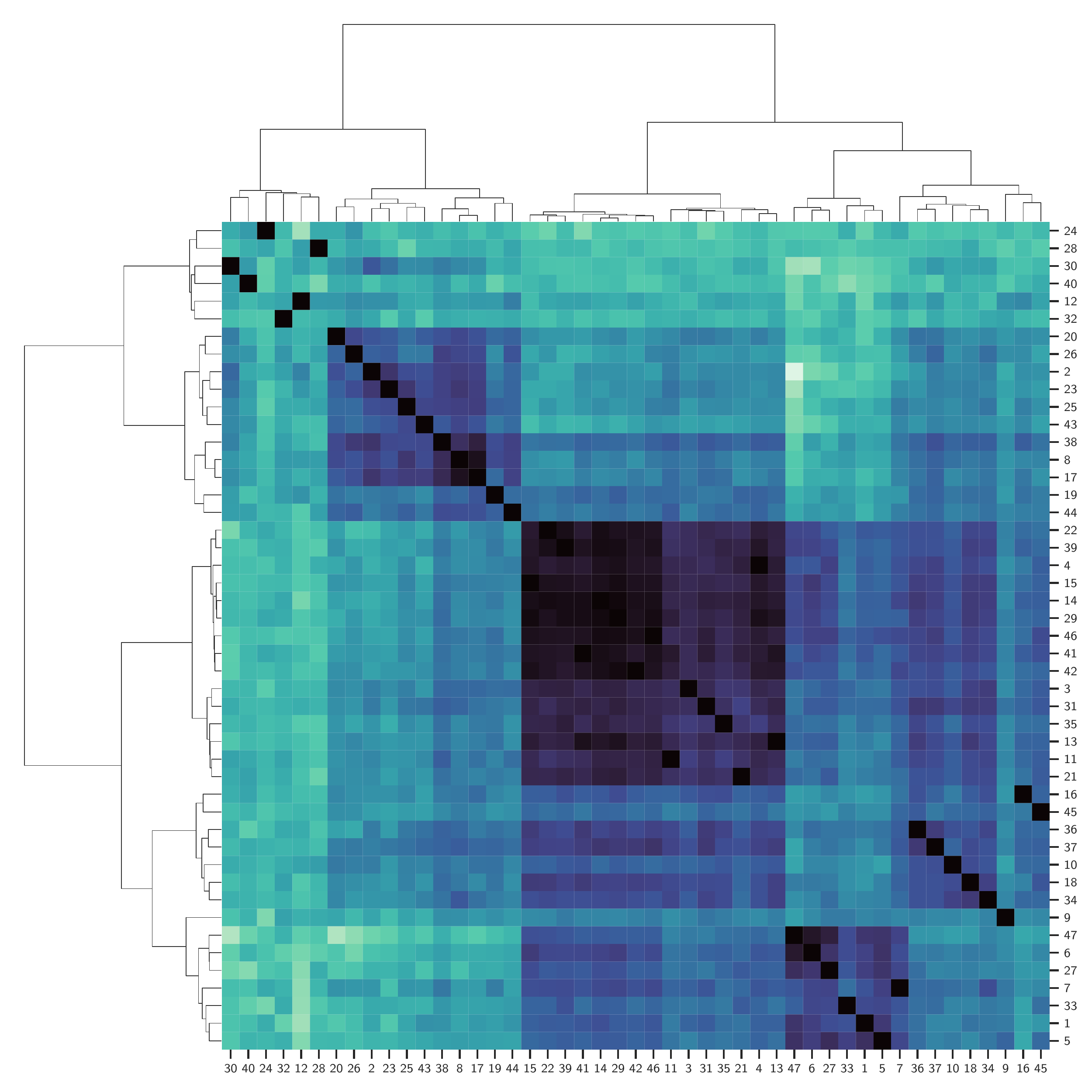}
					\includegraphics[width=0.24\textwidth]{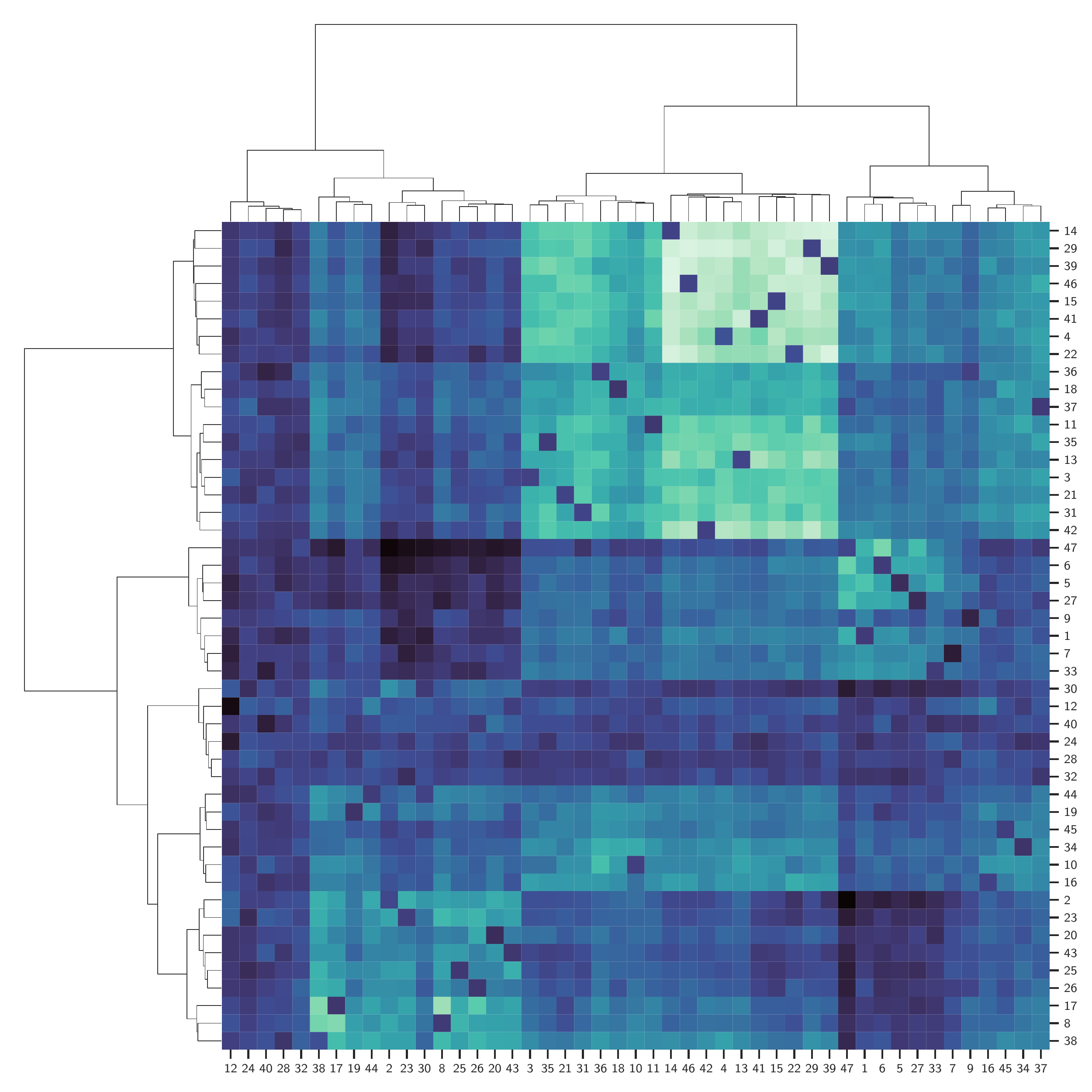}
					\includegraphics[width=0.24\textwidth]{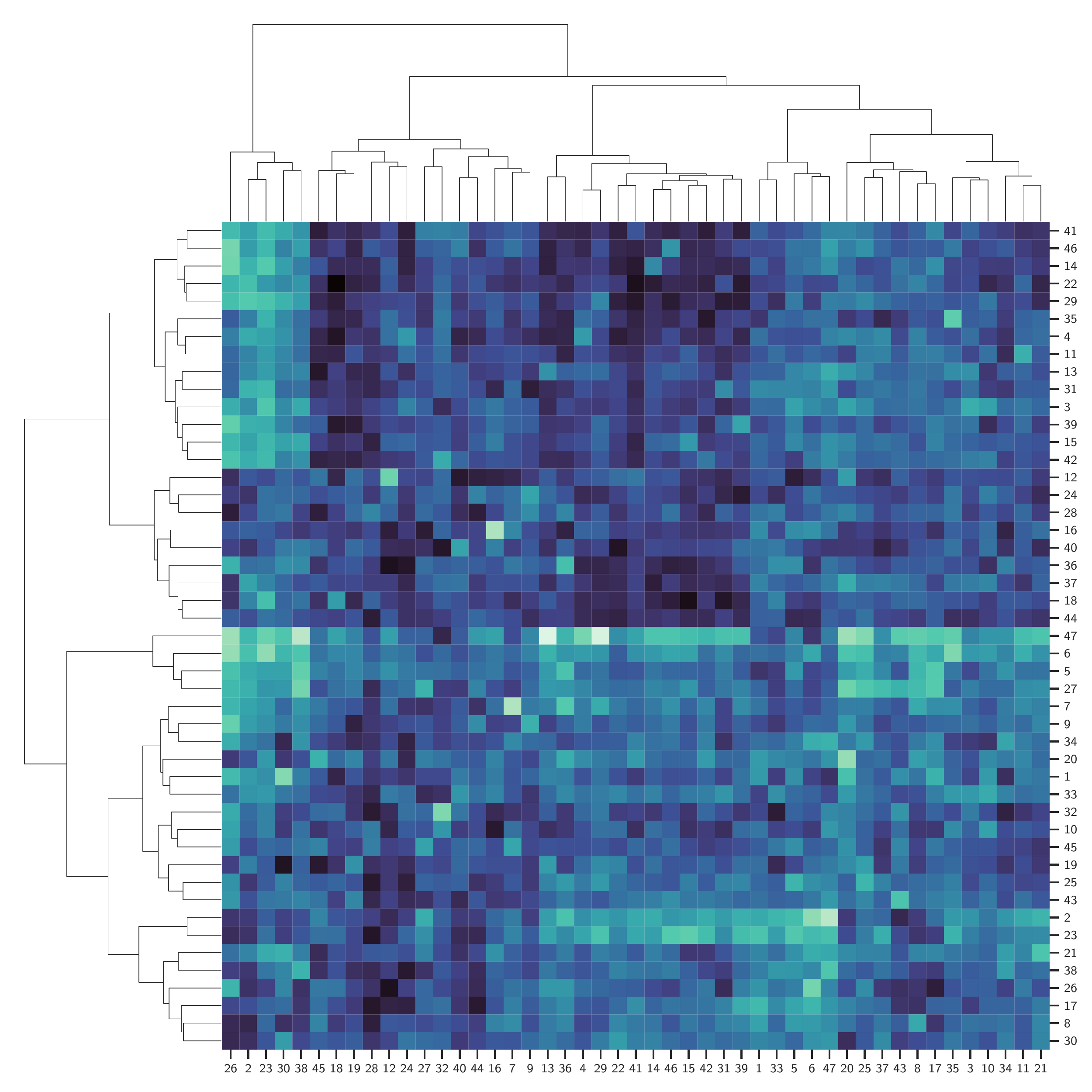}
					\includegraphics[width=0.24\textwidth]{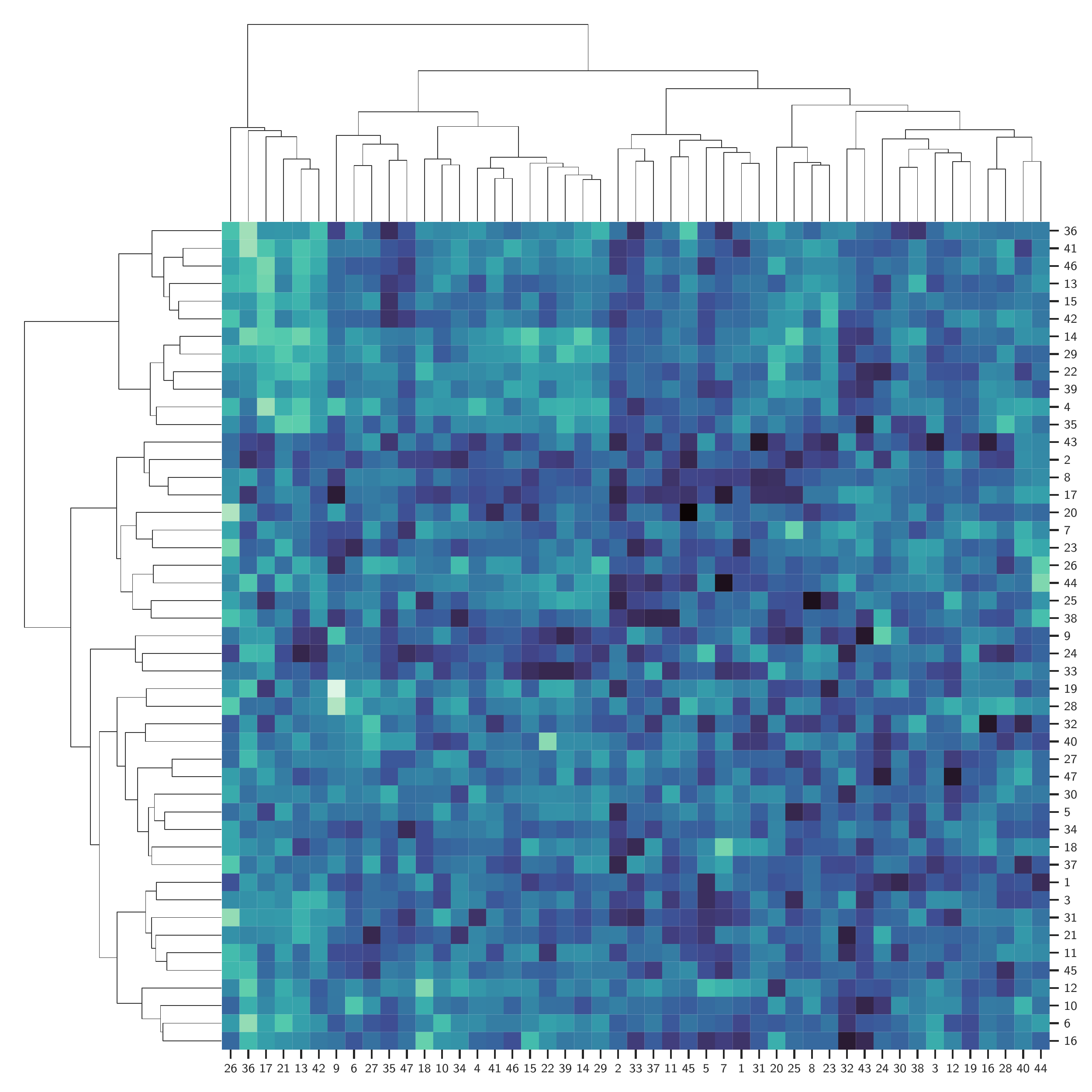}

			\caption{\label{fig:AHC2} AHC implementation on the $\Psi^{\rm pk}(\vartheta\ge 0\sigma_0)$ (upper panels), $\Psi^{\rm up}(\vartheta=0\sigma_0)$ (middle panels), and $\Psi^{\rm pk-tr}(\vartheta_{\rm pk}\ge 0\sigma_0;\vartheta_{\rm tr}\le 0\sigma_0)$ (lower panels). Density plots in each column from left to right correspond to $\tau=0$ Days, $\tau=1$ Day, $\tau=2$ Days, and $\tau=3$ Days, respectively.}
		\end{center}
	\end{figure*}

  To examine the role of time separation in the clustering of excursion sets considered in this paper, we construct the $\Psi^{\rm pk}(\tau;\vartheta)$ for $\tau=0$, $\tau=1$ and $\tau=2$. For each $\tau$, we also consider three thresholds namely,  $\vartheta \ge [-1,0,+1]\sigma_0$. Fig. \ref{fig:AHC2} displays the AHC implementation on the $\Psi(\tau;\vartheta)$. The upper panel corresponds to the  $\Psi^{\rm pk}(\vartheta\ge 0\sigma_0)$ and from left to right, we set $\tau=0$, $\tau =1$, $\tau =2$ and $\tau =3$ days. As we expect, by increasing $\tau$, the magnitude of correlation decreases, and consequently, the results of the AHC method become the same as the random matrix. Also, for $\tau=1$, the value of unweighted TPXF for markets located in the same region decreases, while this value for markets in different regions behaves almost in the opposite way (see Fig. \ref{crossTPCF}). The middle panel of Fig. \ref{fig:AHC2} illustrates  $\Psi^{\rm up}(\vartheta=0\sigma_0)$. The behavior of the mentioned measure is almost the same as peak statistics, however, the robustness of recognizing structures by the AHC method in the context of up-crossing statistics is less than peak statistics. The time dependency of  $\Psi^{\rm pk-tr}(\vartheta_{\rm pk}\ge 0\sigma_0;\vartheta_{\rm tr}\le 0\sigma_0)$ is shown in the lower panel of Fig. \ref{fig:AHC2}. The cross-correlation goes to zero rapidly, leading to a decrease in the capability of structure finding in the AHC approach. We also checked the threshold dependency, and the same results were derived.


		\subsection{Partitioning approach}
		In the previous section, we focused on the clustering notion for the entire time interval, now we evaluate the clustering of excursion sets for the non-overlapping segments. Accordingly, we expect to capture how markets evolve at different time intervals. More precisely, we construct non-overlapping windows with a size $100$ days. Mentioned segmentation constructs 65 sets of time series for each stock market indices. 
		Now, we compute the maximum value of $\xi_{\rm pk-pk}^{\ell \ell'}(i)$ and $\xi_{\rm up-up}^{\ell \ell'}(i)$ at threshold $\vartheta=0\sigma_0$ and finally, we achieve 65 matrices for each mentioned measures. By implementing the singular value decomposition method, the corresponding eigenvalue spectrum is computed for each measure and each segment. Inspired by the similarity measure defined in \cite{Schafer}, we make the following matrix: 
		\begin{equation}
		\Delta_{ij}^{\diamond}\equiv |\lambda_{\rm max}^{\diamond}(i)- \lambda_{\rm max}^{\diamond}(j)| 
		\end{equation}
		where $\diamond$ can be replaced by ``pk-pk" and ``up-up". The $\lambda_{\rm max}$ is the maximum eigenvalue of the matrix constructed for the maximum value of unweighted TPCF of local maxima and up-crossing measures.  Fig. \ref{fig:Peakzeta} illustrates the similarity measures for peak (upper panel) and up-crossing (lower panel).  There are several epochs in these heatmaps that we can observe a drastic difference in similarity with the beginning period used in this research when the global market experienced an almost relatively stable period. According to the prior information given in the literature, one can conclude that observed dissimilarities coincide with well-known financial crises. Precisely, the 1997 Asian financial crisis, 1998 Russian financial crisis, and 1998-1999 Brazilian financial crisis can be noticed. We can also observe the bursting of the dot-com bubble in late 2002. The 2008 Global financial crisis is where we see the highest level of dissimilarity between stable periods, which is evidence of the severity of this crisis compared to other crises that the global market has ever encountered. Going further, we can notice the Euro debt crisis and the 2015-2016 stock market sell-off.
		
	The main achievement of our approach is in the fact that during the periods of stability, the global market has a distinct so-called collective behavior in terms of the clustering of local maxima as well as up-crossing in associated return series. 
	However, this so-called collective behavior changes drastically during the epoch of crisis. We also repeated the above computation for the Pearson correlation coefficient measure, and the same results have been obtained but the distinguishability of existent crises based on excursion sets is almost higher than that of obtained by the Pearson correlation coefficient measure, which is one of advantages of considering excursion sets.

	\section{Summary and Conclusion}\label{conclusion}
	In this study, we employ daily price data from 47 international stock markets and analyze them using the excursion set theory. As the advantage of the number density of famous geometrical measures estimation, theoretically, we first gave a brief explanation of theoretical prediction for a Gaussian stochastic series by using the definition of spectral indices. Going beyond the one-point statistics, the excess probability of finding the pair of critical sets, the so-called clustering, has been defined. To mitigate the boundary effect in computing mentioned clustering measure, three robust estimators have been considered. We also turned to the agglomerative hierarchical clustering procedure and implemented it on the matrices whose elements have been computed based on un-weighted TPCF of excursion sets. To make more sense of the behavior of underlying markets, we took the so-called global approach focusing on the entire time interval of given returns, while on the other hand, the partitioning method utilizes dividing the series into non-overlapping segments. Accordingly, we examined the temporal behavior of markets.  
	
	The number density of local extrema for all data sets used in this study indicated a deviation at the low threshold (almost around the mean value) with respect to the Gaussian theory. This behavior is almost universal for all data sets considered in this paper. This property implies that markets almost experience more ups and downs when the level of returns becomes less than average fluctuation computed for a long period. On the contrary, for the $\vartheta \gtrsim 0$, a deficit number density of local extrema has been recognized (the left panels of Figs. \ref{fig:excursion1}-\ref{fig:excursion4}). Applying the up-crossing condition demonstrated an almost deficiency for the amount of up-crossing at $\vartheta \lesssim 0$ compared to the expected value for the Gaussian series. Based on the dimensional analysis, the inverse of $\langle n_{\rm up}(\vartheta) \rangle$ is proportional to a characteristic time scale for crossing the given threshold, statistically. These behaviors demonstrated the universal property of stock indices studied in this paper. 
	
	The notion of clustering has been utilized in its various meaning in analyzing stochastic fields. Based on the excess probability of finding pairs of the desired feature, we carried out the natural estimator (Eq. (\ref{eq:pp-estimator1})) for computing the un-weighted TPC(X)F of local extrema and up-crossing in stock indices.  By increasing the threshold value, the $\xi_{\rm pk-pk}(\vartheta;\tau)$ and $\xi_{\rm up-up}(\vartheta;\tau)$ grow and statistically achieve their maximum value during one week (Figs. \ref{DualTPCFPeaks1} - \ref{DualTPCFPeaks4}). The unweighted TPC(X)F is almost similar to the random case for $\tau\gtrsim 7$ days. Computing the $\xi_{\rm pk-pk}^{\ell\ell'}$ for $\ell\ne\ell'$ (unweighted TPXF) enables us to assess the statistical synchronization between different markets (Fig. \ref{crossTPCF}). The unweighted TPXF for those markets located in the same geographical region takes its maximum value, while for other pairs depending on the amount of mutual impact, a retarded behavior can be found.

	We also constructed matrices containing $\Psi_{\rm max}^{\rm pk}$, $\Psi_{\rm max}^{\rm up}$, and $\Psi_{\rm max}^{\rm pk-tr}$ for different thresholds and time separations with sizes $47\times 47$. By applying the AHC method on mentioned constructed matrices, we derived the clustering of stock markets when the whole time interval is taken into account (Figs. \ref{fig:AHC1} and \ref{fig:AHC2}).  Our results indicated in Fig. \ref{fig:AHC1} recognized some clusters considerably influenced by geographical region. Interestingly, the peak statistics was more powerful in identifying the different groups. The time separation impact on clustering based on AHC method fed by excursion sets 
	has been examined in  Fig. \ref{fig:AHC2} for $\tau=0,1,2$ days. The best case to recognize clusters happened for $\tau=0$ and the crossing measure is weaker than peak statistics for such purpose when $\tau$ to be increased.

\begin{figure}
		\begin{center}
			\includegraphics[width=0.5\textwidth]{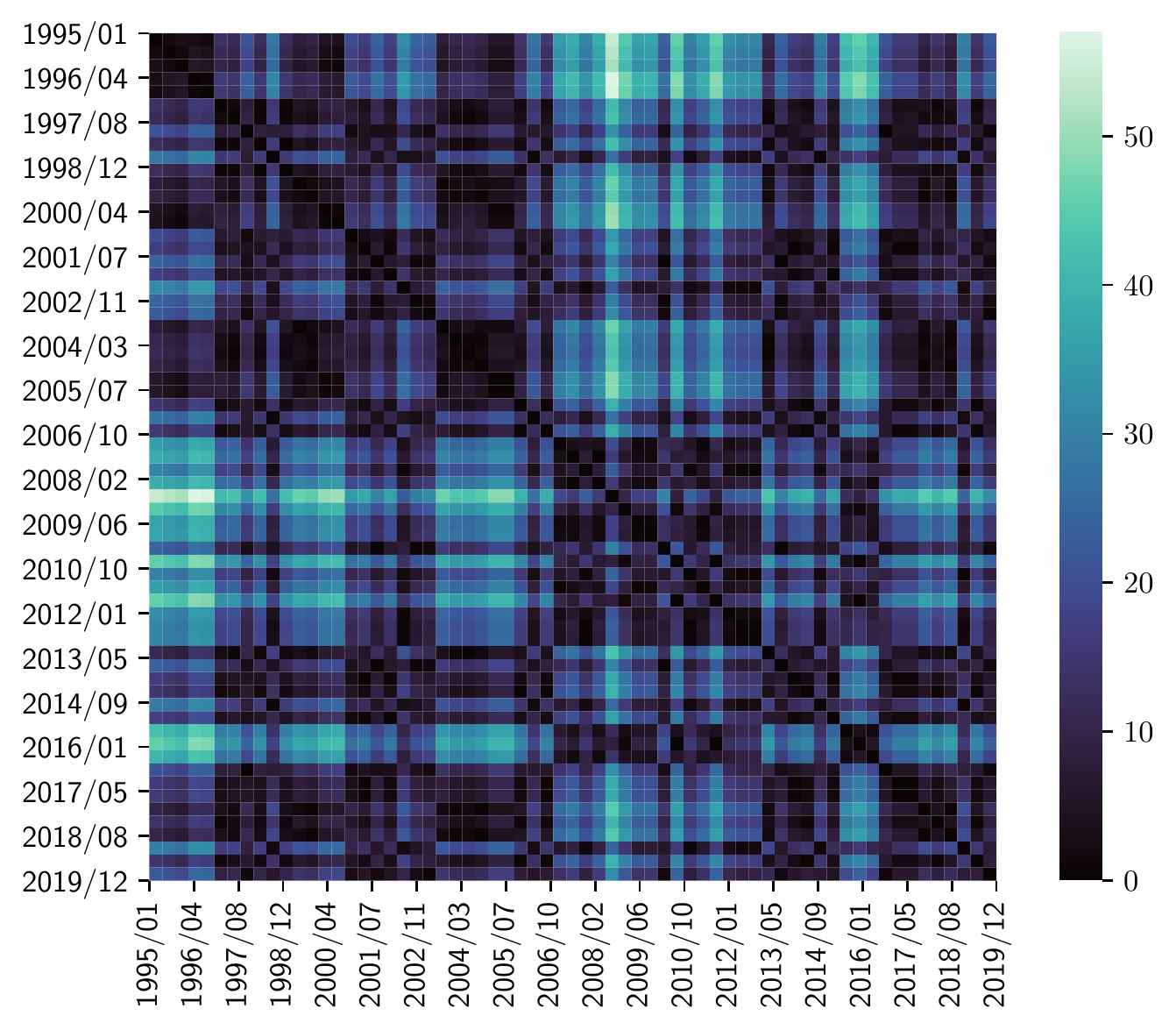}
			\includegraphics[width=0.5\textwidth]{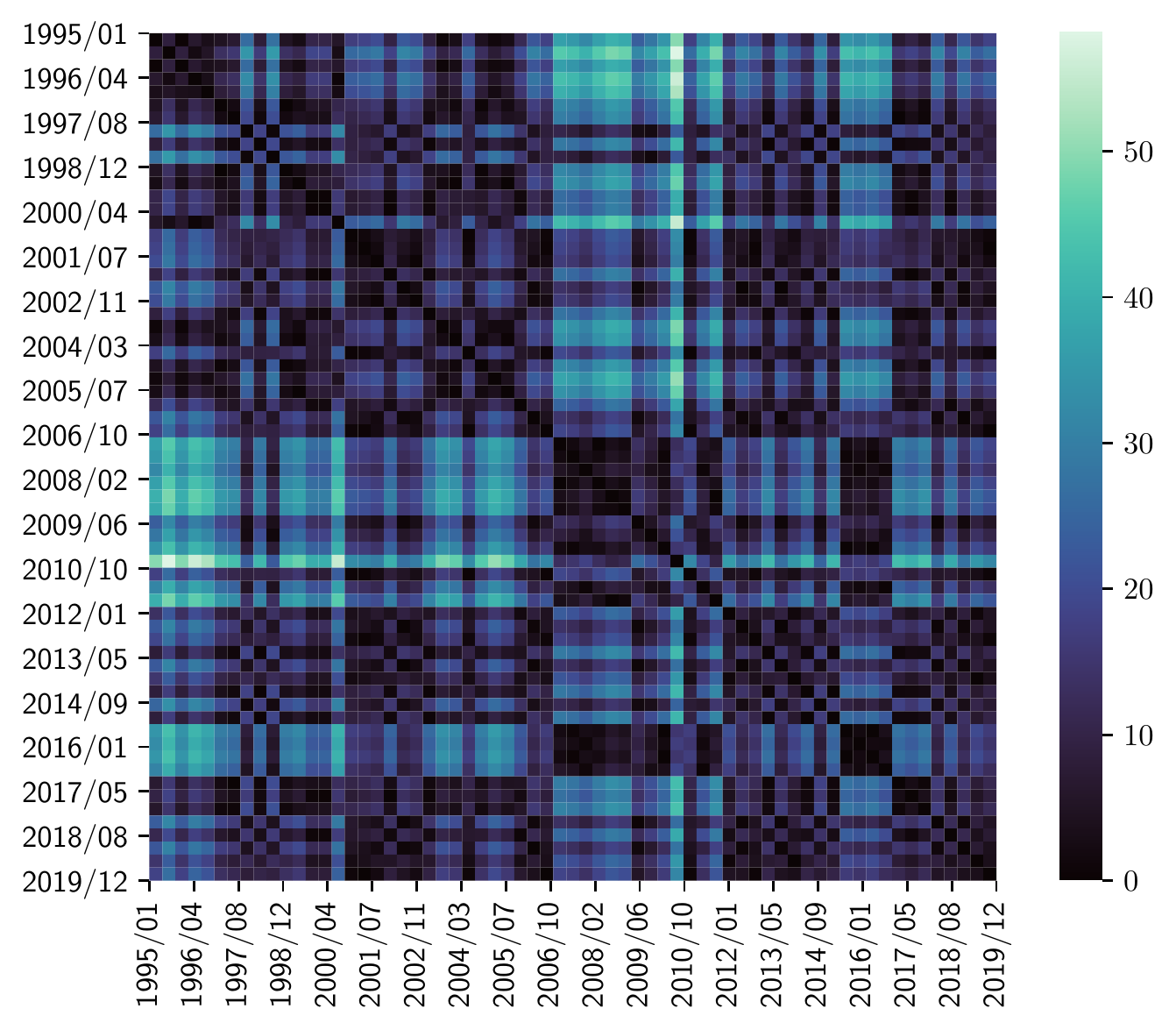}
			\caption{\label{fig:Peakzeta} The matrix representation of $|\lambda^{\diamond}_{\rm max}(i)-\lambda^{\diamond}_{\rm max}(j)|$. The $\lambda^{\diamond}$ is the eigenvalue of the maximum value of unweighted TPCF of peak and up-crossing at threshold $\vartheta=0\sigma_0$ computed for all available data sets. The upper panel corresponds to peak and the lower panel is associated with up-crossing.}
		\end{center}
	\end{figure}

	For the partitioning analysis, we divided our data set into the non-overlapping windows and repeated the calculation of excess probability for each resulting window. Using a similarity measure, we investigated the changes in the correlation structure of the international market and looked for differences between times of stability and crisis. Our results revealed the intense effect of geographical region on how markets behave (i.e., the co-occurrence of critical points). In other words, the	partitioning approach enables us to achieve a better understanding of the market dynamics. The overall structure of the unweighted TPC(X)F of markets experiencing a crisis varies radically compared to the periods of stability. This means, in addition to the drastic and observable changes in market value, a fundamental change occurs in the market dynamics, which is nearly impossible to detect without such sophisticated statistics  (Fig. \ref{fig:Peakzeta}).\\

Final remarks are that utilizing the topological data analysis under the banner of the homology group \citep{zomorodian2005topology11,carlsson2009topology} in addition to excursion sets \citep{Matsubara:2003yt} enables us to make our statistical evaluation of Stock markets more complete and we will leave them for future studies.

\section*{References}


\begin{thebibliography}{10}
\expandafter\ifx\csname url\endcsname\relax
  \def\url#1{\texttt{#1}}\fi
\expandafter\ifx\csname urlprefix\endcsname\relax\def\urlprefix{URL }\fi
\expandafter\ifx\csname href\endcsname\relax
  \def\href#1#2{#2} \def\path#1{#1}\fi

\bibitem{mantegna1999hierarchical}
R.~N. Mantegna, Hierarchical structure in financial markets, The European
  Physical Journal B-Condensed Matter and Complex Systems 11~(1) (1999)
  193--197.

\bibitem{mantegna1999introduction}
R.~N. Mantegna, H.~E. Stanley, Introduction to econophysics: correlations and
  complexity in finance, Cambridge university press, 1999.

\bibitem{voit2005statistical}
J.~Voit, The statistical mechanics of financial markets, Springer Science \&
  Business Media, 2005.

\bibitem{christoffersen2012elements}
P.~Christoffersen, Elements of financial risk management, Academic press, 2012.

\bibitem{chakraborti2011econophysics}
A.~Chakraborti, I.~M. Toke, M.~Patriarca, F.~Abergel, Econophysics review: I.
  empirical facts, Quantitative Finance 11~(7) (2011) 991--1012.

\bibitem{ferreira2017assessment}
P.~Ferreira, A.~Dion{\'\i}sio, S.~Movahed, Assessment of 48 stock markets using
  adaptive multifractal approach, Physica A: Statistical Mechanics and its
  Applications 486 (2017) 730--750.

\bibitem{pagan1996econometrics}
A.~Pagan, The econometrics of financial markets, Journal of empirical finance
  3~(1) (1996) 15--102.

\bibitem{marti2021review}
G.~Marti, F.~Nielsen, M.~Bi{\'n}kowski, P.~Donnat, A review of two decades of
  correlations, hierarchies, networks and clustering in financial markets,
  Progress in Information Geometry (2021) 245--274.

\bibitem{wen2016forecasting}
F.~Wen, X.~Gong, S.~Cai, Forecasting the volatility of crude oil futures using
  har-type models with structural breaks, Energy Economics 59 (2016) 400--413.

\bibitem{gong2017forecasting}
X.~Gong, B.~Lin, Forecasting the good and bad uncertainties of crude oil prices
  using a har framework, Energy Economics 67 (2017) 315--327.

\bibitem{gong2018incremental}
X.~Gong, B.~Lin, The incremental information content of investor fear gauge for
  volatility forecasting in the crude oil futures market, Energy Economics 74
  (2018) 370--386.

\bibitem{gong2018structural}
X.~Gong, B.~Lin, Structural breaks and volatility forecasting in the copper
  futures market, Journal of Futures Markets 38~(3) (2018) 290--339.

\bibitem{he2019risk}
Z.~He, L.~He, F.~Wen, Risk compensation and market returns: The role of
  investor sentiment in the stock market, Emerging Markets Finance and Trade
  55~(3) (2019) 704--718.

\bibitem{yang2020systemic}
X.~Yang, S.~Wen, X.~Zhao, C.~Huang, Systemic importance of financial
  institutions: A complex network perspective, Physica A: Statistical Mechanics
  and its Applications 545 (2020) 123448.

\bibitem{Drozdz}
S.~Dro{\.z}d{\.z}, F.~Gr{\"u}mmer, F.~Ruf, J.~Speth, Towards identifying the
  world stock market cross-correlations: Dax versus dow jones, Physica A:
  Statistical Mechanics and its Applications 294~(1-2) (2001) 226--234.

\bibitem{mantegna}
R.~N. Mantegna, Hierarchical structure in financial markets, The European
  Physical Journal B-Condensed Matter and Complex Systems 11~(1) (1999)
  193--197.

\bibitem{cross}
M.~Eryi{\u{g}}it, R.~Eryi{\u{g}}it, Network structure of cross-correlations
  among the world market indices, Physica A: Statistical Mechanics and its
  Applications 388~(17) (2009) 3551--3562.

\bibitem{diebold2014network}
F.~X. Diebold, K.~Y{\i}lmaz, On the network topology of variance
  decompositions: Measuring the connectedness of financial firms, Journal of
  econometrics 182~(1) (2014) 119--134.

\bibitem{Chakra}
A.~Chakraborty, S.~Easwaran, S.~Sinha, Uncovering the hierarchical structure of
  the international forex market by using similarity metric between the
  fluctuation distributions of currencies, arXiv preprint arXiv:2005.02482.

\bibitem{structure}
J.-B. Zhang, Y.-C. Gao, S.-M. Cai, The hierarchical structure of stock market
  in times of global financial crisis, Physica A: Statistical Mechanics and its
  Applications 542 (2020) 123452.

\bibitem{Schafer}
M.~C. M{\"u}nnix, T.~Shimada, R.~Sch{\"a}fer, F.~Leyvraz, T.~H. Seligman,
  T.~Guhr, H.~E. Stanley, Identifying states of a financial market, Scientific
  reports 2 (2012) 644.

\bibitem{Asian}
H.~Jang, W.~Sul, The asian financial crisis and the co-movement of asian stock
  markets, Journal of Asian Economics 13~(1) (2002) 94--104.

\bibitem{cool}
T.~F. Cooley, V.~Quadrini, Financial markets and firm dynamics, American
  economic review 91~(5) (2001) 1286--1310.

\bibitem{Italo}
L.~S. Junior, I.~D.~P. Franca, Correlation of financial markets in times of
  crisis, Physica A: Statistical Mechanics and its Applications 391~(1-2)
  (2012) 187--208.

\bibitem{RMT1}
E.~P. Wigner, \href{http://www.jstor.org/stable/1970079}{Characteristic vectors
  of bordered matrices with infinite dimensions}, Annals of Mathematics 62~(3)
  (1955) 548--564.
\newline\urlprefix\url{http://www.jstor.org/stable/1970079}

\bibitem{RMT2}
E.~P. Wigner, \href{http://www.jstor.org/stable/1970008}{On the distribution of
  the roots of certain symmetric matrices}, Annals of Mathematics 67~(2) (1958)
  325--327.
\newline\urlprefix\url{http://www.jstor.org/stable/1970008}

\bibitem{Jin}
Y.~Yuan, X.-t. Zhuang, X.~Jin, Measuring multifractality of stock price
  fluctuation using multifractal detrended fluctuation analysis, Physica A:
  Statistical Mechanics and its Applications 388~(11) (2009) 2189--2197.

\bibitem{liao2005clustering}
T.~W. Liao, Clustering of time series data-a survey, Pattern recognition
  38~(11) (2005) 1857--1874.

\bibitem{wu2013hierarchical}
X.-Y. Wu, Z.-G. Zheng, Hierarchical cluster-tendency analysis of the group
  structure in the foreign exchange market, Frontiers of Physics 8~(4) (2013)
  451--460.

\bibitem{lahmiri2016clustering}
S.~Lahmiri, Clustering of casablanca stock market based on hurst exponent
  estimates, Physica A: Statistical Mechanics and its Applications 456 (2016)
  310--318.

\bibitem{jiang2008cluster}
J.~Jiang, W.~Li, X.~Cai, Cluster behavior of a simple model in financial
  markets, Physica A: Statistical Mechanics and its Applications 387~(2-3)
  (2008) 528--536.

\bibitem{lahmiri2017clustering}
S.~Lahmiri, G.~S. Uddin, S.~Bekiros, Clustering of short and long-term
  co-movements in international financial and commodity markets in wavelet
  domain, Physica A: statistical mechanics and its applications 486 (2017)
  947--955.

\bibitem{huang2009comparison}
F.~Huang, P.~Gao, Y.~Wang, Comparison of prim and kruskal on shanghai and
  shenzhen 300 index hierarchical structure tree, in: 2009 International
  Conference on Web Information Systems and Mining, IEEE, 2009, pp. 237--241.

\bibitem{zhao2018stock}
L.~Zhao, G.-J. Wang, M.~Wang, W.~Bao, W.~Li, H.~E. Stanley, Stock market as
  temporal network, Physica A: Statistical Mechanics and its Applications 506
  (2018) 1104--1112.

\bibitem{billio2012econometric}
M.~Billio, M.~Getmansky, A.~W. Lo, L.~Pelizzon, Econometric measures of
  connectedness and systemic risk in the finance and insurance sectors, Journal
  of financial economics 104~(3) (2012) 535--559.

\bibitem{kenett2010dominating}
D.~Y. Kenett, M.~Tumminello, A.~Madi, G.~Gur-Gershgoren, R.~N. Mantegna,
  E.~Ben-Jacob, Dominating clasp of the financial sector revealed by partial
  correlation analysis of the stock market, PloS one 5~(12) (2010) e15032.

\bibitem{su2011non}
C.-W. Su, Non-linear causality between the stock and real estate markets of
  western european countries: Evidence from rank tests, Economic Modelling
  28~(3) (2011) 845--851.

\bibitem{wang2014dynamics}
G.-J. Wang, C.~Xie, P.~Zhang, F.~Han, S.~Chen, Dynamics of foreign exchange
  networks: a time-varying copula approach, Discrete Dynamics in Nature and
  Society 2014.

\bibitem{fiedor2014information}
P.~Fiedor, Information-theoretic approach to lead-lag effect on financial
  markets, The European Physical Journal B 87~(8) (2014) 1--9.

\bibitem{rocchi2017emerging}
J.~Rocchi, E.~Y.~L. Tsui, D.~Saad, Emerging interdependence between stock
  values during financial crashes, Plos one 12~(5) (2017) e0176764.

\bibitem{kenett2010dynamics}
D.~Y. Kenett, Y.~Shapira, A.~Madi, S.~Bransburg-Zabary, G.~Gur-Gershgoren,
  E.~Ben-Jacob, Dynamics of stock market correlations., AUCO Czech Economic
  Review 4~(3).

\bibitem{fiedor2014networks}
P.~Fiedor, Networks in financial markets based on the mutual information rate,
  Physical Review E 89~(5) (2014) 052801.

\bibitem{baitinger2017interconnectedness}
E.~Baitinger, J.~Papenbrock, Interconnectedness risk and active portfolio
  management: the information-theoretic perspective, Available at SSRN 2909839.

\bibitem{barbi2019nonlinear}
A.~Barbi, G.~Prataviera, Nonlinear dependencies on brazilian equity network
  from mutual information minimum spanning trees, Physica A: Statistical
  Mechanics and its Applications 523 (2019) 876--885.

\bibitem{goh2018inference}
Y.~K. Goh, H.~M. Hasim, C.~G. Antonopoulos, Inference of financial networks
  using the normalised mutual information rate, PloS one 13~(2) (2018)
  e0192160.

\bibitem{guo2018development}
X.~Guo, H.~Zhang, T.~Tian, Development of stock correlation networks using
  mutual information and financial big data, PloS one 13~(4) (2018) e0195941.

\bibitem{marti2016optimal}
G.~Marti, F.~Nielsen, P.~Donnat, Optimal copula transport for clustering
  multivariate time series, in: 2016 IEEE International Conference on
  Acoustics, Speech and Signal Processing (ICASSP), IEEE, 2016, pp. 2379--2383.

\bibitem{durante2015cluster}
F.~Durante, R.~Pappada, Cluster analysis of time series via kendall
  distribution, in: Strengthening Links Between Data Analysis and Soft
  Computing, Springer, 2015, pp. 209--216.

\bibitem{brechmann2013hierarchical}
E.~C. Brechmann, Hierarchical kendall copulas and the modeling of systemic and
  operational risk, Ph.D. thesis, Technische Universit{\"a}t M{\"u}nchen
  (2013).

\bibitem{lohre2020hierarchical}
H.~Lohre, C.~Rother, K.~A. Sch{\"a}fer, Hierarchical risk parity: Accounting
  for tail dependencies in multi-asset multi-factor allocations, Machine
  Learning for Asset Management: New Developments and Financial Applications
  (2020) 329--368.

\bibitem{adler81}
R.~Adler, The geometry of random fields, The Geometry of Random Fields,
  Chichester: Wiley, 1981.

\bibitem{adler2011topological}
R.~Adler, J.~E. Taylor, Topological Complexity of Smooth Random Functions:
  {\'E}cole D'{\'E}t{\'e} de Probabilit{\'e}s de Saint-Flour XXXIX-2009,
  Springer Science \& Business Media, 2011.

\bibitem{adler2010persistent}
R.~J. Adler, O.~Bobrowski, M.~S. Borman, E.~Subag, S.~Weinberger, et~al.,
  Persistent homology for random fields and complexes, in: Borrowing strength:
  theory powering applications--a Festschrift for Lawrence D. Brown, Institute
  of Mathematical Statistics, 2010, pp. 124--143.

\bibitem{rice44a}
S.~O. Rice, Mathematical analysis of random noise, Bell Labs Technical Journal
  23~(3) (1944) 282--332.

\bibitem{rice44b}
S.~O. Rice, Mathematical analysis of random noise, The Bell System Technical
  Journal 24~(1) (1945) 46--156.

\bibitem{Bardeen:1985tr}
J.~M. Bardeen, J.~R. Bond, N.~Kaiser, A.~S. Szalay, {The Statistics of Peaks of
  Gaussian Random Fields}, Astrophys. J. 304 (1986) 15--61.
\newblock \href {http://dx.doi.org/10.1086/164143} {\path{doi:10.1086/164143}}.

\bibitem{Bond:1987ub}
J.~Bond, G.~Efstathiou, The statistics of cosmic background radiation
  fluctuations, Monthly Notices of the Royal Astronomical Society 226~(3)
  (1987) 655--687.

\bibitem{ryden1988}
B.~S. Ryden, The area of isodensity contours as a measure of large-scale
  structure, The Astrophysical Journal 333 (1988) L41--L44.

\bibitem{ryd89}
B.~S. Ryden, A.~L. Melott, D.~A. Craig, J.~R. Gott~III, D.~H. Weinberg, R.~J.
  Scherrer, S.~P. Bhavsar, J.~M. Miller, The area of isodensity contours in
  cosmological models and galaxy surveys, The Astrophysical Journal 340 (1989)
  647--660.

\bibitem{mat96a}
T.~Matsubara, Statistics of isodensity contours in redshift space, The
  Astrophysical Journal 457 (1996) 13.

\bibitem{percy00}
P.~H. Brill, A brief outline of the level crossing method in stochastic models,
  CORS Bulletin 34~(4) (2000) 9--21.

\bibitem{Matsubara:2003yt}
T.~Matsubara, Statistics of smoothed cosmic fields in perturbation theory. i.
  formulation and useful formulae in second-order perturbation theory, The
  Astrophysical Journal 584~(1) (2003) 1.

\bibitem{tabar03}
F.~Shahbazi, S.~Sobhanian, M.~R.~R. Tabar, S.~Khorram, G.~Frootan, H.~Zahed,
  Level crossing analysis of growing surfaces, Journal of Physics A:
  Mathematical and General 36~(10) (2003) 2517.

\bibitem{sadegh11}
M.~S. Movahed, S.~Khosravi, Level crossing analysis of cosmic microwave
  background radiation: a method for detecting cosmic strings, Journal of
  Cosmology and Astroparticle Physics 2011~(03) (2011) 012.

\bibitem{sadegh15}
M.~Ghasemi~Nezhadhaghighi, S.~Movahed, T.~Yasseri, S.~M. Vaez~Allaei,
  Characterization of the anisotropy of rough surfaces: Crossing statistics,
  Journal of Applied Physics 122~(8) (2017) 085302.

\bibitem{hadwiger2013vorlesungen}
H.~Hadwiger, Vorlesungen {\"u}ber inhalt, Oberfl{\"a}che und isoperimetrie,
  Vol.~93, Springer-Verlag, 2013.

\bibitem{vafaei2021clustering}
A.~Vafaei~Sadr, S.~Movahed, Clustering of local extrema in planck cmb maps,
  Monthly Notices of the Royal Astronomical Society 503~(1) (2021) 815--829.

\bibitem{matsubara}
T.~Matsubara, Statistics of smoothed cosmic fields in perturbation theory. i.
  formulation and useful formulae in second-order perturbation theory, The
  Astrophysical Journal 584~(1) (2003) 1.

\bibitem{young2001reproductive}
W.~R. Young, A.~J. Roberts, G.~Stuhne, Reproductive pair correlations and the
  clustering of organisms, Nature 412~(6844) (2001) 328--331.

\bibitem{peeb80}
P.~J.~E. Peebles, The large-scale structure of the universe, Princeton
  university press, 1980.

\bibitem{kaiser1984spatial}
N.~Kaiser, On the spatial correlations of abell clusters, The Astrophysical
  Journal 284 (1984) L9--L12.

\bibitem{peac85}
J.~Peacock, A.~F. Heavens, The statistics of maxima in primordial density
  perturbations, Monthly Notices of the Royal Astronomical Society 217~(4)
  (1985) 805--820.

\bibitem{lumusden89}
S.~Lumsden, A.~Heavens, J.~Peacock, The clustering of peaks in a random
  gaussian field, Monthly Notices of the Royal Astronomical Society 238~(2)
  (1989) 293--318.

\bibitem{davis_peeb83}
M.~Davis, P.~Peebles, A survey of galaxy redshifts. v-the two-point position
  and velocity correlations, The Astrophysical Journal 267 (1983) 465--482.

\bibitem{hamilton1993toward}
A.~Hamilton, Toward better ways to measure the galaxy correlation function, The
  Astrophysical Journal 417 (1993) 19.

\bibitem{szaouti98}
I.~Szapudi, A.~S. Szalay, A new class of estimators for the n-point
  correlations, The Astrophysical Journal Letters 494~(1) (1998) L41.

\bibitem{hewet82}
P.~C. Hewett, The estimation of galaxy angular correlation functions, Monthly
  Notices of the Royal Astronomical Society 201~(4) (1982) 867--883.

\bibitem{landy93}
S.~D. Landy, A.~S. Szalay, Bias and variance of angular correlation functions,
  The Astrophysical Journal 412 (1993) 64--71.

\bibitem{Marcos-Caballero:2015lxp}
A.~Marcos-Caballero, R.~Fern\'andez-Cobos, E.~Mart\'\i{}nez-Gonz\'alez,
  P.~Vielva, {The shape of CMB temperature and polarization peaks on the
  sphere}, JCAP 04 (2016) 058.
\newblock \href {http://arxiv.org/abs/1512.07412} {\path{arXiv:1512.07412}},
  \href {http://dx.doi.org/10.1088/1475-7516/2016/04/058}
  {\path{doi:10.1088/1475-7516/2016/04/058}}.

\bibitem{lumsden1989clustering}
S.~Lumsden, A.~Heavens, J.~Peacock, The clustering of peaks in a random
  gaussian field, Monthly Notices of the Royal Astronomical Society 238~(2)
  (1989) 293--318.

\bibitem{rodriguez2014clustering}
A.~Rodriguez, A.~Laio, Clustering by fast search and find of density peaks,
  Science 344~(6191) (2014) 1492--1496.

\bibitem{arabie1996clustering}
P.~Arabie, G.~De~Soete, Clustering and classification, World Scientific, 1996.

\bibitem{tabar2019analysis}
R.~Tabar, Analysis and data-based reconstruction of complex nonlinear dynamical
  systems, Springer, 2019.

\bibitem{matsubara2020statistics}
T.~Matsubara, Statistics of peaks of weakly non-gaussian random fields: Effects
  of bispectrum in two-and three-dimensions, Physical Review D 101~(4) (2020)
  043532.

\bibitem{mayer1940mayer}
J.~Mayer, M.~Goeppert, Mayer, statistical mechanics, JohnW iley \& Sons, New
  York.

\bibitem{pathria}
J.~Rogel-Salazar,
  \href{https://doi.org/10.1080/00107514.2011.603434}{Statistical mechanics,
  3rd edn., by r.k. pathria and p.d. beale}, Contemporary Physics 52~(6) (2011)
  619--620.
\newblock \href
  {http://arxiv.org/abs/https://doi.org/10.1080/00107514.2011.603434}
  {\path{arXiv:https://doi.org/10.1080/00107514.2011.603434}}, \href
  {http://dx.doi.org/10.1080/00107514.2011.603434}
  {\path{doi:10.1080/00107514.2011.603434}}.
\newline\urlprefix\url{https://doi.org/10.1080/00107514.2011.603434}

\bibitem{Landy:1993yu}
S.~D. Landy, A.~S. Szalay, {Bias and variance of angular correlation
  functions}, Astrophys. J. 412 (1993) 64.
\newblock \href {http://dx.doi.org/10.1086/172900} {\path{doi:10.1086/172900}}.

\bibitem{clusterreview}
R.~Xu, D.~Wunsch, Survey of clustering algorithms, IEEE Transactions on neural
  networks 16~(3) (2005) 645--678.

\bibitem{ward}
J.~H. Ward~Jr, Hierarchical grouping to optimize an objective function, Journal
  of the American statistical association 58~(301) (1963) 236--244.

\bibitem{Mullner}
D.~M{\"u}llner, Modern hierarchical, agglomerative clustering algorithms, arXiv
  preprint arXiv:1109.2378.

\bibitem{LW}
G.~N. Lance, W.~T. Williams, A general theory of classificatory sorting
  strategies: 1. hierarchical systems, The computer journal 9~(4) (1967)
  373--380.

\bibitem{jafari2006level}
G.~Jafari, S.~Movahed, S.~Fazeli, M.~R.~R. Tabar, S.~Masoudi, Level crossing
  analysis of the stock markets, Journal of Statistical Mechanics: Theory and
  Experiment 2006~(06) (2006) P06008.

\bibitem{zomorodian2005topology11}
A.~J. Zomorodian, Topology for computing, Cambridge university press, 2005.

\bibitem{carlsson2009topology}
G.~Carlsson, Topology and data, Bulletin of the American Mathematical Society
  46~(2) (2009) 255--308.

\end{thebibliography}

\end{document}